\documentclass[10pt]{article}

\pdfoutput=1

\usepackage[T1]{fontenc}
\usepackage[utf8]{inputenc}

\usepackage{geometry}
\usepackage{titling}
\usepackage[x11names]{xcolor}

\usepackage{amssymb}
\usepackage{dsfont}
\usepackage{bm}
\usepackage{fontawesome}

\usepackage[english]{babel}

\usepackage{graphicx}
\usepackage{subcaption}
\usepackage{booktabs}

\usepackage{amsmath, amsthm}
\usepackage{commath}  
\usepackage[version=4]{mhchem}  
\usepackage{siunitx}

\usepackage{hyperxmp}  
\usepackage[unicode]{hyperref}  
\usepackage{doi}  

	\hypersetup{
		hypertexnames=false,
	}

\usepackage{cite}
\usepackage{chapterbib}

\usepackage{setspace}  
\usepackage[pagewise]{lineno}
\usepackage{todonotes}

\usepackage{catchfile}  
\usepackage{printlen}

\uselengthunit{mm}  

\newcommand{\sep}{, }  

\newcommand{\suppinfo}{Supplementary information}

\newcommand{\eqn}{Equation}
\newcommand{\eqns}{Equations}
\newcommand{\fig}{Figure}
\newcommand{\figs}{Figures}
\newcommand{\tab}{Table}

\newcommand{\startsuppinfo}{
	\pretitle{\begin{center}\Large \suppinfo: \LARGE \par}
	\posttitle{\par\end{center}\vskip 0.5em}

	\setcounter{page}{1}
	\renewcommand*{\thepage}{S\arabic{page}}  

	\setcounter{section}{0}
	\renewcommand*{\thesection}{S\arabic{section}}  
	\renewcommand*{\theHsection}{\thesection}  

	\setcounter{figure}{0}
	\renewcommand*{\thefigure}{S\arabic{figure}}  
	\renewcommand*{\theHfigure}{\thefigure}  

	\setcounter{table}{0}
	\renewcommand*{\thetable}{S\arabic{table}}  
	\renewcommand*{\theHtable}{\thetable}  

	\setcounter{equation}{0}
	\renewcommand{\theequation}{S\arabic{equation}}  
	\renewcommand{\theHequation}{\theequation}  

}

\newcommand{\orcid}[2][]{
	\href{https://orcid.org/#2}{\includegraphics[height=2ex]{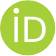}#1}
}
\newcommand{\arxiv}[2][]{
	\href{https://arxiv.org/a/#2}{\includegraphics[height=2ex]{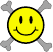}#1}
}
\newcommand{\citex}[2]{\cite[\expandafter{#1}]{#2}}
\newcommand{\dlmf}[3][]{\href{http://dlmf.nist.gov/#2#1#3}{#2#3}}

\newcommand{\wikipedia}[2][en]{%
	\href{https://#1.wikipedia.org/wiki/#2}{#2}%
	\textsuperscript{\faWikipediaW}
}

\DeclareMathOperator{\e}{e}
\DeclareMathOperator{\sech}{sech}

\DeclareMathOperator{\arcsn}{arcsn}
\DeclareMathOperator{\sn}{sn}
\DeclareMathOperator{\cn}{cn}
\DeclareMathOperator{\dn}{dn}
\DeclareMathOperator{\sd}{sd}
\DeclareMathOperator{\cd}{cd}
\DeclareMathOperator{\nd}{nd}

\hypersetup{
	bookmarksopen=true,
	bookmarksnumbered=true,
	linktocpage=true,
}

\newcommand{\documenttitle}{
	Analytical solution for two-dimensional Laplace's equation
	in a shallow domain containing coplanar interdigitated boundaries
}

\newcommand{\authorigivennames}{Cristian F.}
\newcommand{\authorifamilynames}{Guajardo Yévenes}
\newcommand{\authoriorcid}{0000-0001-5595-917X}
\newcommand{\authoriarxiv}{0000-0001-5595-917X}
\newcommand{\authoriemail}{cristian.gua@kmutt.ac.th}
\newcommand{\authoriaffitiation}{
	Biological Engineering Program, Faculty of Engineering
	\\ Pilot Plant Development and Training Institute
	\\ \textsc{King Mongkut's University of Technology Thonburi, Thailand}
}

\newcommand{\authoriigivennames}{Werasak}
\newcommand{\authoriifamilynames}{Surareungchai}
\newcommand{\authoriiorcid}{0000-0002-3639-7279}
\newcommand{\authoriiemail}{werasak.sur@kmutt.ac.th}
\newcommand{\authoriiaffiliation}{
	School of Bioresources and Technology
	\\ Nanoscience \& Nanotechnology Graduate Program
	\\ \textsc{King Mongkut's University of Technology Thonburi, Thailand}
}

\title{\documenttitle}

\author{
	\authorigivennames\ \textsc{\authorifamilynames}
	\orcid{\authoriorcid}\arxiv{\authoriarxiv}
	\\ \texttt{\authoriemail}
	\\ \authoriaffitiation
	\and
	\authoriigivennames\ \textsc{\authoriifamilynames}
	\orcid{\authoriiorcid}
	\\ \texttt{\authoriiemail}
	\\ \authoriiaffiliation
}

\date{January 7, 2021}

\hypersetup{
	pdftitle={\documenttitle},
	pdfsubject={Laplace's equation appears frequently in physical applications involving conservable quantities.
Among these applications, miniaturized devices have been of interest, in particular those using interdigitated arrays.
Therefore, we solved the two-dimensional Laplace's equation for a shallow or finite domain consisting of interdigitated boundaries.
We achieved this by using Jacobian elliptic functions to conformally transform the interdigitated domain into a parallel plates domain.
The obtained expressions for potential distribution, flux density and flux allow for arbitrary domain height, different band widths and asymmetric potentials at the interdigitated array, besides considering fringing effects at both ends of the array.
All these expressions depend only on relative dimensions, instead of absolute ones.
With these results we showed that the behavior in shallow or finite domains approaches that of a semi-infinite domain, when its height is greater than the separation between the centers of consecutive bands.
We also found that, for any desired but fixed flux, bands of equal width minimize the total surface of the interdigitated array.
Finally, we present approximate expressions for the flux, based on elementary functions, which can be applied to ease the calculation of currents (faradaic or non-faradaic), capacitances and resistances among other possible applications.
},
	pdfkeywords={Interdigitated arrays\sep
Finite domain\sep
Shallow domain\sep
Steady state\sep
Laplace's equation\sep
Schwarz-Christoffel transformations\sep
Jacobian elliptic functions},
	pdfauthor={
		\authorigivennames\ \authorifamilynames,
		\authoriigivennames\ \authoriifamilynames
	},
}

\begin{document}


\maketitle

\section*{Abstract}
Laplace's equation appears frequently in physical applications involving conservable quantities.
Among these applications, miniaturized devices have been of interest, in particular those using interdigitated arrays.
Therefore, we solved the two-dimensional Laplace's equation for a shallow or finite domain consisting of interdigitated boundaries.
We achieved this by using Jacobian elliptic functions to conformally transform the interdigitated domain into a parallel plates domain.
The obtained expressions for potential distribution, flux density and flux allow for arbitrary domain height, different band widths and asymmetric potentials at the interdigitated array, besides considering fringing effects at both ends of the array.
All these expressions depend only on relative dimensions, instead of absolute ones.
With these results we showed that the behavior in shallow or finite domains approaches that of a semi-infinite domain, when its height is greater than the separation between the centers of consecutive bands.
We also found that, for any desired but fixed flux, bands of equal width minimize the total surface of the interdigitated array.
Finally, we present approximate expressions for the flux, based on elementary functions, which can be applied to ease the calculation of currents (faradaic or non-faradaic), capacitances and resistances among other possible applications.

\clearpage
\section{Introduction}


Several physical applications involve the \emph{density} $\rho_{Q}$ of some \emph{conservable quantity} $Q$ (and a related potential), for example:
concentration (which is also a potential in the mathematical sense),
heat density (temperature), or electric charge density (electric potential).
See \tab\ \ref{main:tab:quantities} for details.
If there is a \emph{flux density}\footnote{%
	In vector calculus, the \emph{flux density} $\bm{j}$ corresponds to the amount of quantity flowing per unit area per unit time,
	while the \emph{flux} $i$ corresponds to the amount of quantity flowing per unit time.
	In transport theory, in contrast, $\bm{j}$ is called \emph{flux} while $i$ is called \emph{flow rate}.
	See Wikipedia's entry \wikipedia{Flux}.
}
$\bm{j}$ modifying the amount of $Q$ inside the domain of interest,
then the \emph{continuity equation} applies.
In particular, if such flux density is proportional only to the gradient of some \emph{potential} $u$,
then the continuity equation is reduced to
\begin{equation}
	\label{main:eqn:diffusion}
	\dpd{\rho_{Q}}{t} = -\bm{\nabla} \cdot \bm{j}
	\quad \text{and} \quad
	\bm{j} = -\gamma \bm{\nabla} u
	\quad \Rightarrow \quad
	\dpd{\rho_{Q}}{t} = \gamma \nabla^{2} u
\end{equation}
In these cases, \emph{Laplace's equation} appears naturally
when the density of the conservable quantity inside the domain of interest is zero ($\rho_{Q} = 0$) or,
more commonly, when this density doesn't change in time ($\partial \rho_{Q}/\partial t = 0$)
\begin{equation}
	\label{main:eqn:laplace}
	\nabla^{2} u = 0
\end{equation}


Among real applications, miniaturized (and more recently wearable) electronics, actuators and sensors have been of increasing interest.
In particular, devices using \emph{interdigitated arrays} (IDA) are a popular choice \cite{Mamishev2004may,Atalay2018may,BroselOliu2019dec}, due to ease of fabrication and advantages in performance \cite{Szunerits2007,ElKady2013feb,Miserere2015}.
This motivates the use of Laplace's equation in a finite or shallow domain with interdigitated boundaries.

\begin{table}
	\centering
	\begin{tabular}{rlrlrl}
	\toprule
	\multicolumn{2}{c}{Density of quantity}
	& \multicolumn{2}{c}{Flux density of quantity}
	& \multicolumn{2}{c}{Auxiliary variables}
	\\ \midrule
	$\rho_{Q}$ &
	& $\bm{j} = -\gamma \bm{\nabla} u$ &
	& $u$ &
	\\
	$c$ & \si{\mole\per\metre\cubed}
	& $\bm{\varphi} = -D\bm{\nabla} c$ & \si{\mole\per\metre\squared\per\second}
	& -- &
	\\
	$\rho_{H} = c \rho_{M} T$ & \si{\joule\per\metre\cubed}
	& $\bm{q} = -k \bm{\nabla} T$ & \si{\joule\per\metre\squared\per\second}
	& $T$ & \si{\kelvin}
	\\
	$\rho_{E} = \bm{\nabla} \cdot \bm{D}$ & \si{\coulomb\per\metre\cubed}
	& $\bm{j} = -\sigma \bm{\nabla} V$ & \si{\coulomb\per\metre\squared\per\second}
	& $\bm{D} = -\epsilon \bm{\nabla} V$ & \si{\coulomb\per\metre\squared}
	\\ \bottomrule
\end{tabular}%

	\caption{
		\emph{Density of conservable quantity}, flux density, related potentials or other auxiliary variables, and their proportionality constants.
		\emph{Generic density} $\rho_{Q}$, generic flux density $\bm{j}$, generic potential $u$ and generic proportionality constant $\gamma$.
		\emph{Concentration} $c$, diffusion flux (density) $\bm{\varphi}$, diffusion coefficient $D$.
		\emph{Heat density} $\rho_{H}$, heat flux (density) $\bm{q}$, temperature $T$, specific heat capacity $c$, mass density $\rho_{M}$ and thermal conductivity $k$.
		\emph{Electric charge density} $\rho_{E}$, electric current density $\bm{j}$, electric potential $V$, electric displacement $\bm{D}$, electric conductivity $\sigma$ and absolute permittivity~$\epsilon$. 
	}
	\label{main:tab:quantities}
\end{table}


Solving this equation enables us to know the distribution of the potential $u$ in steady state, which allows the computation of the flux density $\bm{j}$ and subsequently the flux $i$.
An analytical expression for the latter is important in real applications,
since the flux of a quantity can be measured experimentally
when applying a difference of potentials at the IDA.
Some examples of flux include: faradaic current (in electrochemistry), rate of heat flow, and non-faradaic current (for electronic devices).
The latter can be directly related to capacitantes and resistances.


Analytically solving Laplace's equation inside a finite or shallow domain with interdigitated boundaries is challenging due to its mixed boundary conditions:
Potentials at the IDA bands (Dirichlet boundaries) alternate with insulations at the gaps (Neumann boundaries).
Several attempts have been made to solve this problem, among which we would like to highlight important results obtained for electrochemistry and resistive/capacite sensors.


In electrochemistry, Aoki and colleagues \cite{Aoki1988dec} were probably the first in explaining, analytically and accurately, the faradaic current generated experimentally at IDA electrodes in a semi-infinite diffusion domain.
Their accurate equation considered bands of different width, and it was obtained by solving Laplace's equation through conformal transformations (elliptic functions).
Additionally, they provided an approximated equation using simple elementary functions, which is easy to evaluate, but it is only valid for wide IDA bands.
Almost two decades later, Morf and coworkers \cite{Morf2006may}, based on the results obtained in \cite{Aoki1988dec}, found another approximated expression which is valid for smaller widths, covering most practical cases.
However, these results for semi-infinite geometries do not always apply for the case of finite ones.
Despite the current use of IDA electrodes in finite or shallow diffusion domains, no analytical expressions have been developed to describe them.


In the case of sensors, Wei \cite{Wei1977apr} found an analytical expression for the capacitance at IDA electrodes in a semi-infinitely thick medium.
His equation was obtained by using elliptic functions as conformal transformations for Laplace's equation, and considered symmetric potentials applied at IDA bands of equal width.
Almost two decades later, Olthuis and colleagues \cite{Olthuis1995mar} explained experimentally measured capacitances in a semi-infinite medium, with similar conformal transformations, but allowing asymmetric potentials applied at bands of equal width.
Finally, a decade later, Igreja and Dias \cite{Igreja2004may} explained experimental capacitances in (multilayer) finite media.
Their equations approach that of semi-infinite medium when this is thick enough, and considered the case of equal band widths and symmetric potentials.
All the previous approaches \cite{Wei1977apr,Olthuis1995mar,Igreja2004may} produced accurate expressions based on elliptic functions.
Moreover, the expressions found in \cite{Igreja2004may} had increased accuracy, since they considered fringing effects at both ends of the IDA.
However, in these cases, there is a loss in flexibility by considering only bands of equal width and symmetric potentials.
Also, no analytical expressions for the potential distribution and flux density were presented, and no approximate expressions using elementary functions were given to ease the calculations of capacitances (flux).


In this work, we derived new conformal transformations to solve Laplace's equation, which consider a finite or shallow domain and different band widths at the IDA.
This led to improved expressions (based on elliptic functions) for the generic potential, flux density and flux, which allow for arbitrary domain height, asymmetric potentials applied at bands of different widths, and consider fringing effects at both ends of the IDA.
With these results, we found that bands of equal width minimize the total surface of the IDA for any desired but fixed flux.
We present new simplified expressions for the flux, in terms of elementary functions, for the case of IDA with different band widths whenever possible.
These approximated expressions allow easy and quick evaluation of the flux.
Finally, we verified under what combinations of band widths and domain height these approximations are valid (error less than \SI{+-5}{\percent} with respect to their ideal counterpart).

\section{Description of the problem}

\begin{figure}
	\centering
	\subcaptionbox{%
		\label{main:fig:domain:full:3d}
		Full view.  
	}{\includegraphics{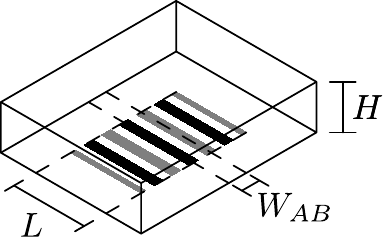}}
	\hfill
	\subcaptionbox{
		\label{main:fig:domain:full:2d}
		Cross section.  
	}{\includegraphics{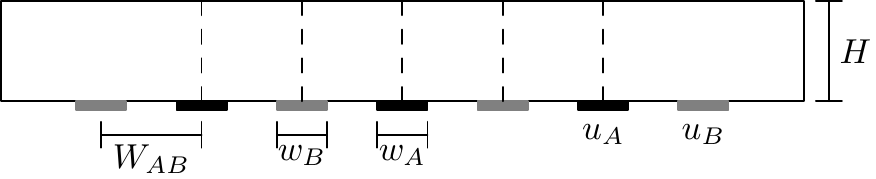}}
	\caption{%
		\emph{Practical} configuration of an interdigitated array (IDA) in a rectangular box of finite height $H$.
		The IDA consists of two arrays of $N_{A}$ and $N_{B}$ bands each (shown $N_{A} + 1 = N_{B} = 4$) at which the potentials $u_{A}$ and $u_{B}$ are applied.
		The two-dimensional domain can be approximated as an assembly of several \emph{interior cells} (four shown) and two \emph{exterior cells} at the ends of the IDA.
	}
	\label{main:fig:domain:full}
\end{figure}

Let an IDA be located at the bottom of a rectangular box of height $H$, whose walls behave as perfect insulators (\fig\ \ref{main:fig:domain:full}).
The IDA consists of two arrays, $A$ (black) and $B$ (gray), of $N_{A}$ and $N_{B}$ bands each, where two consecutive bands are separated by a distance $W_{AB}$ from each of their centers.
The bands have width of $w_{A}$ and $w_{B}$, a common length $L$, and are assumed to have no elevation (or thickness) above the bottom of the domain.

For the rectangular box we will consider two cases.
In an \emph{ideal} case, the IDA fits perfectly inside the domain, such that all four vertical walls touch the borders of the IDA, and the two most exterior bands (one at each end) have only half width, which results in $N_{A} = N_{B}$.
In a more \emph{practical} configuration (\fig\ \ref{main:fig:domain:full:3d}), the vertical walls don't touch the IDA, and the most exterior bands have full width.
In this last case, we will consider $N_{A} + 1 = N_{B}$ in order to take advantage of the symmetry produced at both ends of the IDA.
However, the results can be extrapolated to the case $N_{A} = N_{B}$ without much difficulty if desired.

The problem consists of finding the potential distribution $u$ inside the rectangular box, and the flux density $\bm{j}$ and flux $i$ at the IDA, when we 
apply potentials $u_{A}$ and $u_{B}$ at the arrays $A$ and $B$ respectively.

\section{Methods}

\subsection{Reduction to an assembly of two-dimensional cells}

If the IDA fits exactly within an \emph{ideal} domain, the rectangular box can be reduced to two dimensions, due to the symmetry along the bands.
This reduced domain can be further modeled as an assembly of two-dimensional cells, due to the vertical symmetry boundaries at the center of each band.

For a more \emph{practical} case, the domain can be reduced to two dimensions (\fig\ \ref{main:fig:domain:full:2d}) when the bands are sufficiently long, which ensures that fringing effects at the ends of each band can be neglected.
This can be further reduced to an assembly of two-dimensional cells, if the number of bands is sufficiently large, which ensures that fringing effects at both ends of the IDA can also be neglected.
In this case, the symmetry boundaries between cells remain vertical at the interior of the IDA.
However, they will bend near the ends of the IDA, where our approximation of vertical symmetry boundaries will produce some error.
Choosing the last vertical symmetry boundary to be located one band before each end of the IDA helps to better account for bendings of this boundary at the last band.
Regarding the bands, they are allowed to have a thickness or elevation different from zero, as long as it is sufficiently small compared with their widths.

Thus the IDA in a \emph{practical domain} consists of an assembly of several \emph{interior cells} and two \emph{exterior cells} at the ends of the IDA
(\fig\ \ref{main:fig:domain:full:2d}), and the solution of Laplace's equation will hold accurately.
Whereas, the IDA in an \emph{ideal domain} consists only of an assembly of \emph{interior cells}, and the solution of Laplace's equation will be exact.

\subsection{Conformal transformation of a generic cell}
\label{main:sec:T}

In practice, directly solving Laplace's equation for the \emph{interior} and \emph{exterior cells} is complicated.
This is due to alternated boundaries of potential and insulation at the bottom of the cells.
However, by using domain transformations, it is possible to rearrange these boundary conditions, so they can be placed at different walls in a transformed cell.
Complex conformal transformations are useful, since they leave Laplace's equation (as well as potential and insulation/symmetry boundaries) invariant under domain changes \cite[\S5.7]{Driscoll2002} \cite[\S6]{Olver2020jun}.
In particular, \emph{Jacobian elliptic functions} are of interest, since they conformally map a square domain into the upper half-plane \cite[\S2.5]{Driscoll2002} \citex{\S\dlmf{22.}{18.ii}}{dlmf}.
Also the \emph{Möbius functions} are important, since they are able to reorganize the upper half-plane, by mapping into itself \cite[\S2.3]{Driscoll2002}.

\begin{figure}
	\centering
	\includegraphics{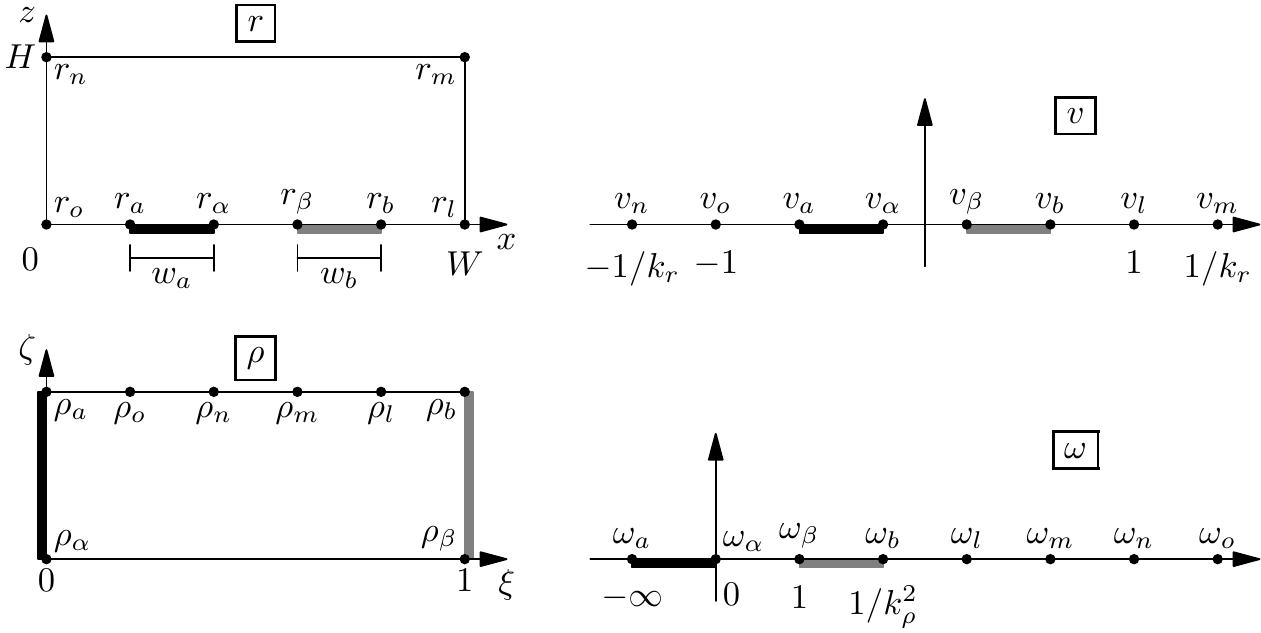}
	\caption{%
		Complex transformation of the \emph{generic cell} of IDA domain $\bm{r} = (x, z)$ into the conformal parallel-plates domain $\bm{\rho} = (\xi, \zeta)$, by using the auxiliary complex domains $\bm{v}$ and $\bm{\omega}$.
		First, the transformation $\bm{r} \to \bm{v}$ maps the interior of the cell into the upper half-plane by using a Jacobian elliptic function.
		Later, the transformation $\bm{v} \to \bm{\omega}$ reorganizes the structure of the upper half-plane through a Möbius transformation.
		Finally, the transformation $\bm{\omega} \to \bm{\rho}$ maps the upper half-plane to the interior of the parallel-plates cell, by using a composition of squared root and an inverse Jacobian elliptic function.
	}
	\label{main:fig:transformation:generic}
\end{figure}

Let $\bm{r} = (x, z) = x + \bm{i}z$ designate the coordinates of a \emph{generic cell} (\fig\ \ref{main:fig:transformation:generic}) where $\bm{i}^{2} = -1$.
This \emph{generic cell} represents simultaneously \emph{interior} and \emph{exterior cells} of an IDA, depending on how its parameters are chosen.
In case of an \emph{interior cell}, one chooses
\begin{subequations}
	\label{main:eqn:cell:int-ext}
	\begin{equation}
		\label{main:eqn:cell:int}
		\bm{r}_{o} = \bm{r}_{a} = 0
		,\quad w_{a} = w_{A}/2
		,\quad w_{b} = w_{B}/2
		,\quad \bm{r}_{b} = W_{AB} = \bm{r}_{l} = W
	\end{equation}%
	whereas, in case of an \emph{exterior cell}
	\begin{equation}
		\label{main:eqn:cell:ext}
		\bm{r}_{o} = \bm{r}_{a} = 0
		,\quad w_{a} = w_{A}/2
		,\quad w_{b} = w_{B}
		,\quad \bm{r}_{b} = W_{AB} + w_{B}/2 < \bm{r}_{l} = W
	\end{equation}%
\end{subequations}%
This generic domain can be conformally transformed by
\begin{subequations}
	\label{main:eqn:T}
	\begin{equation}
		\bm{\rho} = T_{r}^{\rho}(\bm{r})
		= T_{\omega}^{\rho} \circ T_{v}^{\omega} \circ T_{r}^{v}(\bm{r})
	\end{equation}%
	passing through the auxiliary domains $\bm{v}$ and $\bm{\omega}$
	\begin{align}
		\label{main:eqn:T:r-v}
		\bm{v} = T_{r}^{v}(\bm{r})
		&= -\cd\!\del{ K(k_{r}) \frac{2\bm{r}}{W}, k_{r} }
		\\
		\label{main:eqn:T:v-omega}
		\bm{\omega} = T_{v}^{\omega}(\bm{v})
		&= \frac{(\bm{v}-\bm{v}_{\alpha})}{(\bm{v}-\bm{v}_{a})}
		\frac{(\bm{v}_{\beta}-\bm{v}_{a})}{(\bm{v}_{\beta}-\bm{v}_{\alpha})}
	\end{align}%
	into a parallel-plates domain of coordinates $\bm{\rho} = (\xi, \zeta) = \xi + \bm{i}\zeta$
	\begin{equation}
		\label{main:eqn:T:omega-rho}
		\bm{\rho} = T_{\omega}^{\rho}(\bm{\omega})
		= \frac{1}{K(k_{\rho})} \arcsn(\sqrt{\bm{\omega}}, k_{\rho})
	\end{equation}%
\end{subequations}%
where Laplace's equation has a simple solution.
Here, the moduli and their complements are given by
\begin{subequations}
	\label{main:eqn:T:kr-krho}
	\begin{align}
		\label{main:eqn:T:kr}
		q_{r} = Q(k_{r}) = \exp\del{ -\pi \frac{2H}{W} },
		&\quad
		q_{r}' = Q(k_{r}') = \exp\del{ -\pi \frac{W}{2H} }
		\\
		\label{main:eqn:T:krho}
		k_{\rho}^{2} =
		\frac{
			(\bm{v}_{b} - \bm{v}_{a}) (\bm{v}_{\beta} - \bm{v}_{\alpha})
		}{
			(\bm{v}_{b} - \bm{v}_{\alpha}) (\bm{v}_{\beta} - \bm{v}_{a})
		},
		&\quad
		{k_{\rho}'}^{2} =
		\frac{
			(\bm{v}_{b} - \bm{v}_{\beta}) (\bm{v}_{\alpha} - \bm{v}_{a})
		}{
			(\bm{v}_{b} - \bm{v}_{\alpha}) (\bm{v}_{\beta} - \bm{v}_{a})
		}
	\end{align}%
\end{subequations}%
and the points $\bm{r}_{a}$, $\bm{r}_{\alpha}$, $\bm{r}_{\beta}$ and $\bm{r}_{b}$ on the boundary of the IDA domain $\bm{r}$ are transformed to
the auxiliary domain $\bm{v}$ as
\begin{subequations}
	\label{main:eqn:T:vAB}
	\begin{align}
		\bm{v}_{a} = -\cd\!\del{ K(k_{r})\frac{2 \bm{r}_{a}}{W}, k_{r} },
		& \quad
		\bm{v}_{\alpha}
		= -\cd\!\del{ K(k_{r})\frac{2(\bm{r}_{a} + w_{a})}{W}, k_{r} }
		\\
		\bm{v}_{b} = -\cd\!\del{ K(k_{r})\frac{2 \bm{r}_{b}}{W}, k_{r} },
		& \quad
		\bm{v}_{\beta}
		= -\cd\!\del{ K(k_{r})\, \frac{2(\bm{r}_{b} - w_{b})}{W}, k_{r} }
	\end{align}%
\end{subequations}%
and finally to the parallel-plates domain $\bm{\rho}$ as
\begin{subequations}
	\label{main:eqn:T:rhoAB}
	\begin{align}
		\bm{\rho}_{a}
		= \xi_{a} + \bm{i} \zeta_{a}
		= 0 + \bm{i} \frac{K'(k_{\rho})}{K(k_{\rho})},
		&\quad
		\bm{\rho}_{\alpha}
		= \xi_{\alpha} + \bm{i} \zeta_{\alpha}
		= 0 + \bm{i} 0
		\\
		\bm{\rho}_{b}
		= \xi_{b} + \bm{i} \zeta_{b}
		= 1 + \bm{i} \frac{K'(k_{\rho})}{K(k_{\rho})},
		&\quad
		\bm{\rho}_{\beta}
		= \xi_{\beta} + \bm{i} \zeta_{\beta}
		= 1 + \bm{i} 0
	\end{align}%
\end{subequations}%
For a detailed construction of this transformation see \emph{\suppinfo\ \S\ref{supp:sec:T}}.

Several special functions and definitions have been used for the conformal transformation $T_{r}^{\rho}$:
$\cd(\bm{z},k)$ and $\arcsn(\bm{z},k)$ correspond to \emph{Jacobian elliptic functions} \citex{\eqns\ (\dlmf[E]{22.2.}{8}) and (\dlmf[E]{22.15.}{12})}{dlmf} analogous to their circular counterparts $\cos()$ and $\arcsin()$ respectively,
$K(k)$ and $K'(k)$ correspond to the \emph{complete elliptic integral of the first kind} and its associated function respectively \citex{\eqns\ (\dlmf[E]{19.2.}{4}), (\dlmf[E]{19.2.}{8}) and (\dlmf[E]{19.2.}{9})}{dlmf},
$Q(k)$ corresponds to the \emph{elliptic nome function} \citex{\eqn\ (\dlmf[E]{22.2.}{1})}{dlmf},
and $k$ and $k'=\sqrt{1 - k^{2}}$ correspond to the \emph{elliptic modulus} and its complement, which are used as parameters for elliptic functions.

\section{Results}

\subsection{Solution for the generic cell}
\label{main:sec:solution}

By using the transformation of the last section, Laplace's equation and its boundary conditions (potential and insulation/symmetry) can be written in the parallel-plates domain $\bm{\rho} = (\xi, \zeta)$ as
\begin{subequations}
	\begin{align}
		\dpd[2]{u}{\xi}(\xi, \zeta) + \dpd[2]{u}{\zeta}(\xi, \zeta) = 0 &
		\\
		u(0, \zeta) = u_{A},
		\quad u(1, \zeta) = u_{B},&
		\quad \dpd{u}{\zeta}(\xi, 0) = \dpd{u}{\zeta}(\xi, \zeta_{a}) = 0
	\end{align}%
\end{subequations}%
where $\zeta_{a} = \Im\bm{\rho}_{a} = K'(k_{\rho})/K(k_{\rho})$ is the imaginary part of $\bm{\rho}_{a}$ in \eqns\ \eqref{main:eqn:T:rhoAB}.

\begin{figure}[t]
	\centering
	\subcaptionbox{%
		Tall domain ($H/W_{AB} = 1$).
	}{
		\includegraphics{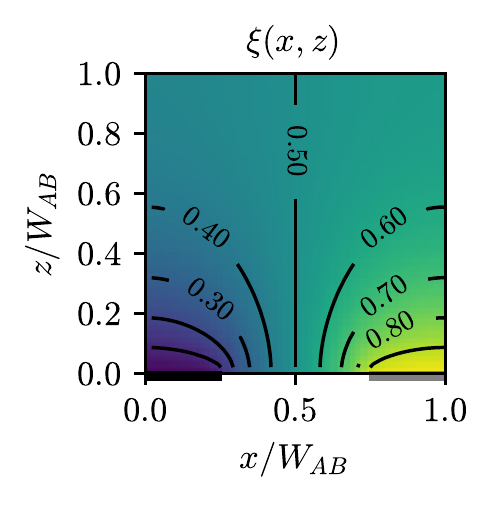}%
		\includegraphics{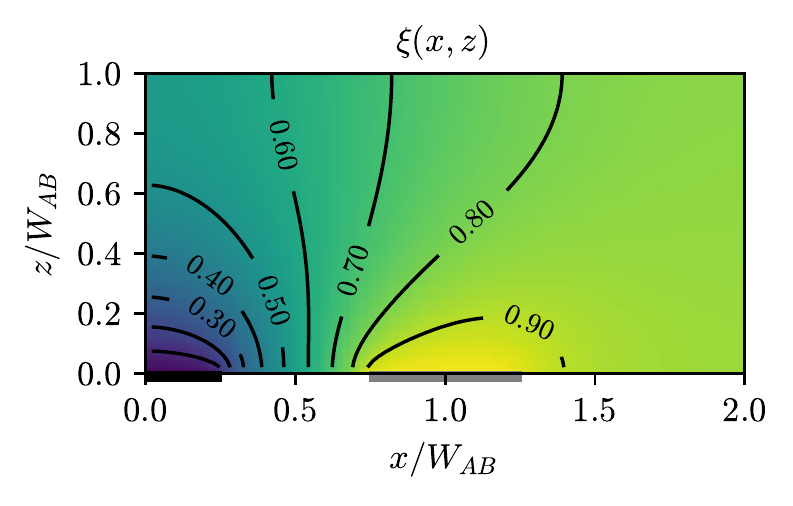}
	}
	\subcaptionbox{%
		Shallow domain ($H/W_{AB} = \num{0,6}$).
	}{
		\includegraphics{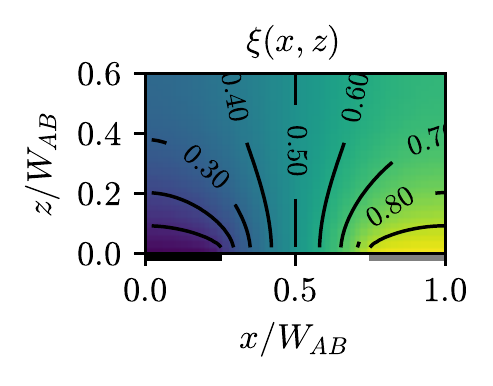}%
		\includegraphics{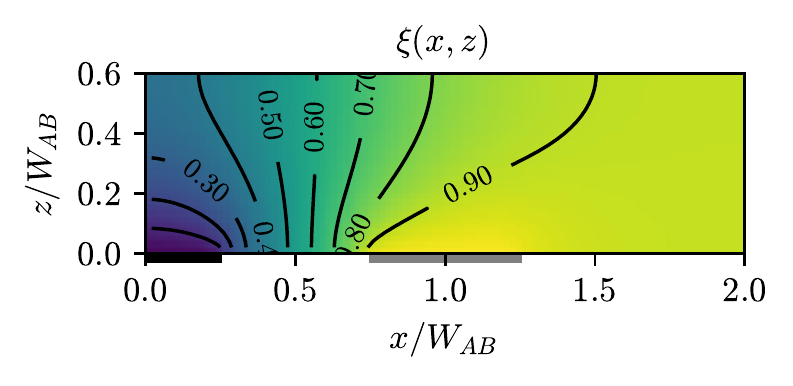}
	}
	\caption{%
		Normalized potential distribution $\xi(x,z) = [u(x,z) - u_{A}]/[u_{B} - u_{A}]$ for equal band widths $w_{A} = w_{B} = \num{0,5} W_{AB}$ and different domain heights $H/W_{AB}$.
		Left and right figures correspond to \emph{interior} ($W = W_{AB}$) and \emph{exterior} ($W = 2W_{AB}$) cells respectively.
	}
	\label{main:fig:potential}
\end{figure}

This equation has a straightforward solution $u(\xi, \zeta) = u_{A} + [u_{B} - u_{A}] \xi$, which corresponds to a linear interpolation of the potentials at each plate.
This result must be carried back to the \emph{generic} IDA domain $\bm{r} = (x, z)$ to obtain the final solution
\begin{equation}
	\label{main:eqn:potential}
	u(x, z) = u_{A} + [u_{B} - u_{A}] \xi(x, z)
\end{equation}
where $\xi(x, z) = \Re T_{r}^{\rho}(x, z)$ corresponds to the real part of the conformal transformation in \emph{Methods \S\ref{main:sec:T}}.
See \fig\ \ref{main:fig:potential} for potential distributions at \emph{interior} and \emph{exterior cells}.

\begin{figure}
	\centering
	\includegraphics{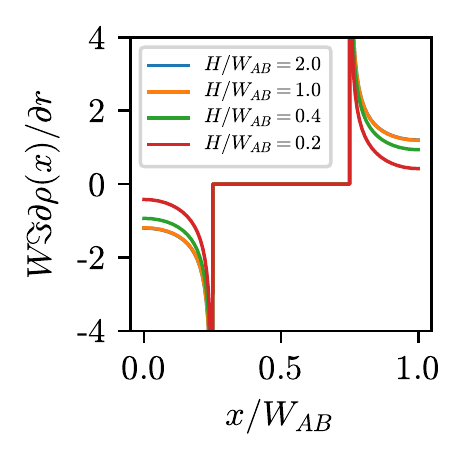}%
	\includegraphics{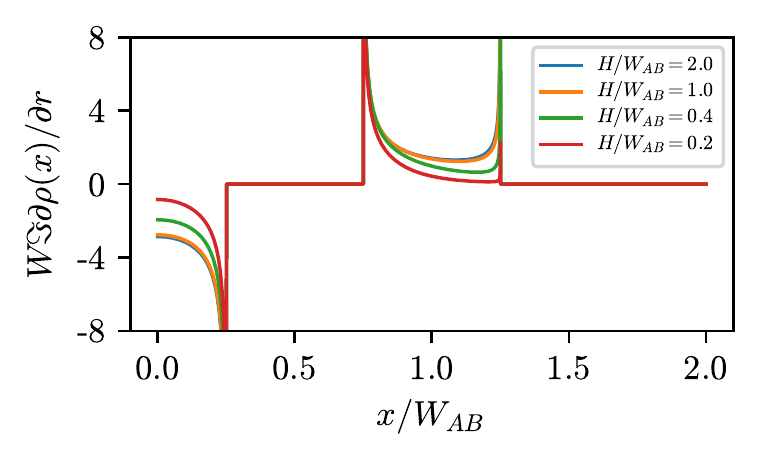}
	\caption{%
		Normalized flux density $W \Im\partial\bm{\rho}(x)/\partial\bm{r} = W j(x)/(\gamma [u_{B} -u_{A}])$ for equal band widths $w_{A} = w_{B} = \num{0,5} W_{AB}$ and different domain heights $H/W_{AB}$.
		Left and right figures correspond to \emph{interior} ($W = W_{AB}$) and \emph{exterior} ($W = 2 W_{AB}$) cells respectively.
	}
	\label{main:fig:flux_density}
\end{figure}

Since the IDA bands are equipotential surfaces, the flux density at the bands must be perpendicular to them, and therefore it must possess only vertical component
\begin{subequations}
	\label{main:eqn:flux_density}
	\begin{equation}
		j(x)
		= -\gamma \dpd{u}{z}(x, 0)
		= -\gamma [u_{B} - u_{A}] \dpd{\xi}{z}(x, 0)
		=  \gamma [u_{B} - u_{A}] \Im \dpd{\bm{\rho}}{\bm{r}}(x)
	\end{equation}%
	where the last equality is due to the Cauchy-Riemann equations \cite[Theorem 3.2]{Olver2020jun}
	\begin{equation}
		\label{main:eqn:cauchy-riemann}
		\dpd{\xi}{z}
		= -\dpd{\zeta}{x}
		= -\Im\dpd{\bm{\rho}}{x}
		= -\Im\dpd{\bm{\rho}}{\bm{r}}
	\end{equation}%
	After some algebraic calculations (see \emph{\suppinfo\ \S\ref{supp:sec:dTdr}}), we obtained the derivative of the conformal transformation $\bm{\rho}(\bm{r}) = T_{r}^{\rho}(\bm{r})$
	\begin{equation}
		\label{main:eqn:dTdr}
		\dpd{\bm{\rho}}{\bm{r}}
		=
		\frac{\bm{i}}{W} \frac{K(k_{r})}{K(k_{\rho})}\,
		\frac{
			(\bm{v}_{b}-\bm{v}_{\alpha})^{1/2}
			(\bm{v}_{\beta}-\bm{v}_{a})^{1/2}
		}{
			(\bm{v}-\bm{v}_{\alpha})^{1/2}
			(\bm{v}-\bm{v}_{\beta})^{1/2}
		}
		\frac{
			(1 - \bm{v}^{2})^{1/2}
			(1 - k_{r}^{2} \bm{v}^{2})^{1/2}
		}{
			(\bm{v}-\bm{v}_{a})^{1/2}
			(\bm{v}_{b}-\bm{v})^{1/2}
		}
	\end{equation}%
	where $\bm{v}$ is given in \eqn\ \eqref{main:eqn:T:r-v} and the parameters are given in \eqns\ \eqref{main:eqn:cell:int-ext}, \eqref{main:eqn:T:kr-krho} and \eqref{main:eqn:T:vAB}.
\end{subequations}%
See \fig\ \ref{main:fig:flux_density} for flux densities at the bottom of \emph{interior} and \emph{exterior cells}.

\begin{figure}
	\centering
	\subcaptionbox{%
		\label{main:fig:flux:interior}
		Interior cell ($W/W_{AB} = 1$).
	}{%
		\includegraphics{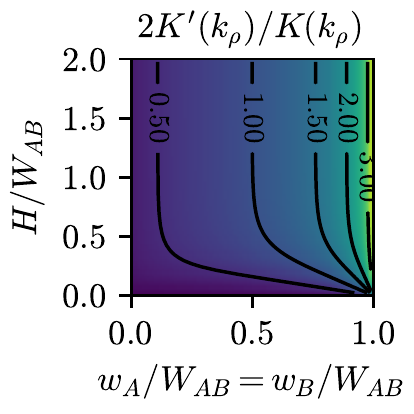}%
		\includegraphics{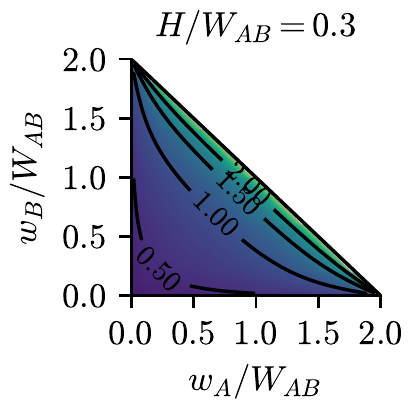}%
		\includegraphics{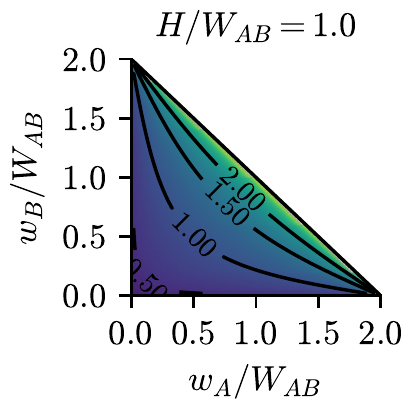}%
	}
	\subcaptionbox{%
		\label{main:fig:flux:exterior}
		Exterior cell ($W/W_{AB} = 2$).
	}{%
		\includegraphics{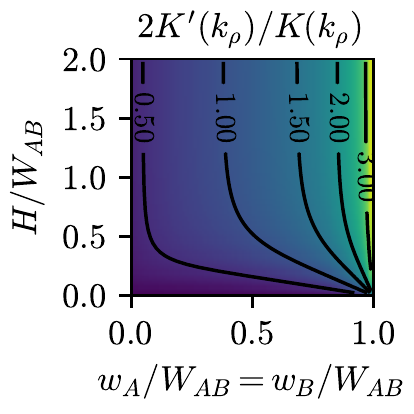}%
		\includegraphics{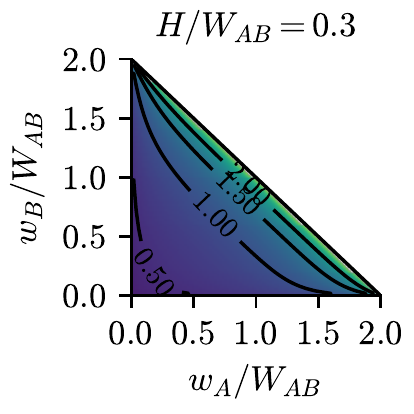}%
		\includegraphics{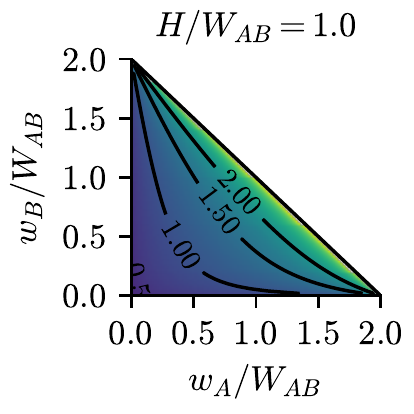}%
	}
	\caption{%
		Selected two-dimensional slices of the normalized flux per band $2K'(k_{\rho})/K(k_{\rho}) = (i_{A}/L)/(\gamma [u_{A} - u_{B}])$ as a function of the cells dimensions.
		Plots considering \emph{exterior cells} with $2 \leq W/W_{AB} \leq 5$ presented almost no variations (shown at \emph{\suppinfo\ \S\ref{supp:sec:flux}}).
	}
	\label{main:fig:flux}
\end{figure}

Finally, we can obtain the flux per band by integrating the flux density on the surface of a band $E \in \{ A,B \}$
\begin{subequations}
	\label{main:eqn:flux:pre}
	\begin{equation}
		i_{E}
		= \int_{E} j(x) L\dif{x}
		= -L \gamma [u_{B} - u_{A}] \int_{E} \dpd{\xi}{z}(x, 0) \dif{x}
	\end{equation}
	which can be simplified using the Cauchy-Riemann identities in \eqn\ \eqref{main:eqn:cauchy-riemann}
	\begin{equation}
		-\int_{E} \dpd{\xi}{z}(x, 0) \dif{x}
		= \Im \int_{E} \dpd{\bm{\rho}}{x}(x, 0) \dif{x}
		= \Im \int_{E} \partial{\bm{\rho}}(\bm{r})
	\end{equation}
	Due to symmetry, we can take twice the integral over half band for the \emph{interior cells}
	\begin{equation}
		\left. \Im \int_{A} \partial{\bm{\rho}}(\bm{r})
		= 2 \Im \bm{\rho}(\bm{r}) \right|_{\bm{r}_{a}}^{\bm{r}_{\alpha}}
		,\quad
		\left. \Im \int_{B} \partial{\bm{\rho}}(\bm{r})
		= 2 \Im \bm{\rho}(\bm{r}) \right|_{\bm{r}_{\beta}}^{\bm{r}_{b}}
	\end{equation}%
	For the \emph{exterior cells}, we just combine both cells to obtain a full band of $A$, while for $B$ only one \emph{exterior cell} suffices
	\begin{equation}
		\left. \Im \int_{A} \partial{\bm{\rho}}(\bm{r})
		= 2 \Im \bm{\rho}(\bm{r}) \right|_{\bm{r}_{a}}^{\bm{r}_{\alpha}}
		,\quad
		\left. \Im \int_{B} \partial{\bm{\rho}}(\bm{r})
		= \Im \bm{\rho}(\bm{r}) \right|_{\bm{r}_{\beta}}^{\bm{r}_{b}}
	\end{equation}%
\end{subequations}%
Combining \eqns\ \eqref{main:eqn:flux:pre} with \eqref{main:eqn:T:rhoAB} we have
\begin{subequations}
	\label{main:eqn:flux:band}
	\begin{equation}
		\label{main:eqn:flux:band:int}
		-i_{A}^\text{int} = i_{B}^\text{int}
		= L\, 2\frac{K'(k_{\rho})}{K(k_{\rho})} \bigg|_\text{int}
		\cdot \gamma [u_{B} - u_{A}]
	\end{equation}
	for bands in the \emph{interior cells}, and
	\begin{equation}
		-i_{A}^\text{ext} = 2 i_{B}^\text{ext}
		= L\, 2\frac{K'(k_{\rho})}{K(k_{\rho})} \bigg|_\text{ext}
		\cdot \gamma [u_{B} - u_{A}]
	\end{equation}%
\end{subequations}%
for bands in the \emph{exterior cells}.
The value of $k_{\rho}$ can be obtained from \eqns\ \eqref{main:eqn:cell:int-ext}, \eqref{main:eqn:T:kr-krho} and \eqref{main:eqn:T:vAB} for \emph{interior} and \emph{exterior cells}.
See \fig\ \ref{main:fig:flux} for flux per band as a function of the dimensions of \emph{interior} and \emph{exterior cells}.

\subsection{Simplified flux for tall and shallow domains}

The flux depends on elliptic functions which are normally evaluated using specialized software.
However, expressions involving elementary functions are desirable for using any general purpose software or calculator.
For this reason, we simplified the expression for the flux by approximating the ratio $K'(k_{\rho})/K(k_{\rho})$.
In particular, we chose to simplify that of an \emph{interior cell}, since this is the only kind of cell in an \emph{ideal domain}, and also the dominant one in case of a more \emph{practical domain} (see \emph{Discussion~\S\ref{main:sec:total_flux}}).

As shown in \cite[\eqn\ (29)]{Aoki1988dec} and \cite[\eqns\ (4) and (5)]{Morf2006may}, this ratio can be approximated for sufficiently small moduli $k_{\rho}$ or $k_{\rho}'$ by
\begin{subequations}
	\label{main:eqn:K'K:approx}
	\begin{align}
		\frac{K'(k_{\rho})}{K(k_{\rho})} \bigg|_{k_{\rho} \approx 0}
		&\approx -\frac{1}{\pi} \ln\del{ \frac{k_{\rho}^{2}}{16} }
		\label{main:eqn:K'K:k0}
		\\
		\frac{K'(k_{\rho})}{K(k_{\rho})} \bigg|_{k_{\rho}' \approx 0}
		&\approx -\pi \sbr{ \ln\del{ \frac{{k_{\rho}'}^{2}}{16} } }^{-1}
		\label{main:eqn:K'K:k'0}
	\end{align}%
\end{subequations}%

In our case, the moduli $k_{\rho}$ and $k_{\rho}'$ in \eqns\ \eqref{main:eqn:cell:int}, \eqref{main:eqn:T:krho} and \eqref{main:eqn:T:vAB} also depend on elliptic functions
\begin{subequations}
	\begin{equation}
		k_{\rho}^{2} = \frac{
			\displaystyle
			2 \sbr{
				\cd\del{ K(k_{r}) \frac{w_{A}}{W_{AB}},\, k_{r} } +
				\cd\del{ K(k_{r}) \frac{w_{B}}{W_{AB}},\, k_{r} }
			}
		}{
			\displaystyle
			\sbr{ 1 + \cd\del{ K(k_{r}) \frac{w_{A}}{W_{AB}},\, k_{r} } }
			\sbr{ 1 + \cd\del{ K(k_{r}) \frac{w_{B}}{W_{AB}},\, k_{r} } }
		}
	\end{equation}%
	\begin{equation}
		{k_{\rho}'}^{2} = \frac{
			\displaystyle
			\sbr{ 1 - \cd\del{ K(k_{r}) \frac{w_{A}}{W_{AB}},\, k_{r} } }
		}{
			\displaystyle
			\sbr{ 1 + \cd\del{ K(k_{r}) \frac{w_{A}}{W_{AB}},\, k_{r} } }
		}
		\frac{
			\displaystyle
			\sbr{ 1 - \cd\del{ K(k_{r}) \frac{w_{B}}{W_{AB}},\, k_{r} } }
		}{
			\displaystyle
			\sbr{ 1 + \cd\del{ K(k_{r}) \frac{w_{B}}{W_{AB}},\, k_{r} } }
		}
	\end{equation}%
\end{subequations}%
and therefore must be approximated too.
To accomplish this, we found alternative representations, which must be easier to approximate than their original counterparts.
The one for $k_{\rho}$ depends on the gap $g = W_{AB} - w_{A} = W_{AB} - w_{B}$ between bands of equal width $w_{A} = w_{B}$
\begin{subequations}
	\begin{equation}
		k_{\rho}^{2} = \frac{
			\displaystyle
			4\, \sn\del{ K(k_{r}) \frac{g}{W_{AB}},\, k_{r} }
		}{
			\displaystyle
			\sbr{ 1 + \sn\del{ K(k_{r}) \frac{g}{W_{AB}},\, k_{r} } }^{2}
		}
	\end{equation}%
	due to quarter period translation $\cd(\bm{u} + K(k),k) = -\sn(\bm{u}, k)$ \citex{\tab\ \dlmf[T]{22.4.}{3}}{dlmf} and doble rotation of the argument $\sn(-\bm{u}, k) = \sn(\bm{i}^{2}\bm{u}, k) = -\sn(\bm{u}, k)$ \citex{\tab\ \dlmf[T]{22.6.}{1}}{dlmf}.
	The one for $k_{\rho}'$ can be expressed using different band widths
	\begin{equation}
		{k_{\rho}'}^{2} = {k_{r}'}^{4} \frac{
			\displaystyle
			\sd\del{ \frac{K(k_{r})}{2} \frac{w_{A}}{W_{AB}},\, k_{r} }^{2}
		}{
			\displaystyle
			\cn\del{ \frac{K(k_{r})}{2} \frac{w_{A}}{W_{AB}},\, k_{r} }^{2}
		}
		\frac{
			\displaystyle
			\sd\del{ \frac{K(k_{r})}{2} \frac{w_{B}}{W_{AB}},\, k_{r} }^{2}
		}{
			\displaystyle
			\cn\del{ \frac{K(k_{r})}{2} \frac{w_{B}}{W_{AB}},\, k_{r} }^{2}
		}
	\end{equation}%
\end{subequations}%
due to $[1 - \cd(2\bm{u},k)]/[1 + \cd(2\bm{u},k)] = {k'}^{2} \sd(\bm{u},k)^{2}/\cn(\bm{u},k)^{2}$ \cite[\eqns\ (1.10) and (4.1)]{Carlson2004nov}.
Here the modulus $k_{r}$ is obtained from $q_{r} = Q(k_{r}) = \exp(-\pi 2H/{W}_{AB})$ in \eqn\ \eqref{main:eqn:T:kr}, and the special functions $\sn$, $\sd$, $\cn$ and $\cd$ correspond to \emph{Jacobian elliptic functions} analogous to their circular counterparts $\sin$ and $\cos$ \citex{\S\dlmf{22.}{2}}{dlmf}.

Finally, we approximated $k_{\rho}$ and $k_{\rho}'$ for the cases of tall and shallow domains.
In case of tall domains ($H$ is large) $q_{r} \to 0^{+}$, $k_{r} \to 0^{+}$ and $k_{r}' \to 1^{-}$, which causes $K(k_{r}) \to \pi/2$, $\sn \to \sin$, $\sd \to \sin$ and $\cd \to \cos$ \citex{\eqn\ (\dlmf[E]{19.6.}{1}) and \tab\ \dlmf[T]{22.5.}{3}}{dlmf} or \cite[\eqns\ (10)]{Fenton1982jul}
\begin{subequations}
	\label{main:eqn:moduli:Hinf}
	\begin{align}
		\label{main:eqn:krho:Hinf}
		\bigg. k_{\rho}^{2} \bigg|_{
			\scriptsize \shortstack{
				$H \to +\infty$ \\
				$g \approx 0$
			}
		}
		&=
		4 \sin\del{ \frac{\pi}{2} \frac{g}{W_{AB}} }
		\sbr{ 1 + \sin\del{ \frac{\pi}{2} \frac{g}{W_{AB}} } }^{-2}
		\hspace{-1em} \approx 4 \del{ \frac{\pi}{2}\frac{g}{W_{AB}} }
		\\
		\label{main:eqn:krho':Hinf}
		\bigg. {k_{\rho}'}^{2} \bigg|_{
			\scriptsize \shortstack{
				$H \to +\infty$ \\
				$w_{A} \approx 0$ \\
				$w_{B} \approx 0$
			}
		}
		&=
		\tan\del{ \frac{\pi}{4} \frac{w_{A}}{W_{AB}} }^{2}
		\tan\del{ \frac{\pi}{4} \frac{w_{B}}{W_{AB}} }^{2}
		\approx
		\del{ \frac{\pi}{4} \frac{w_{A}}{W_{AB}} }^{2}
		\del{ \frac{\pi}{4} \frac{w_{B}}{W_{AB}} }^{2}
	\end{align}%
\end{subequations}%
In case of shallow domains ($H$ is small) $q_{r}' \to 0^{+}$ and $k_{r}' \to 0^{+}$, which causes $\sn \to \tanh$, $\sd \to \sinh$ and $\cn \to \sech$ \citex{\dlmf[T]{22.5.}{4}}{dlmf}.
However, to avoid the divergence in $K(k_{r}) \to +\infty$ \citex{\eqn\ (\dlmf[E]{19.6.}{1})}{dlmf}, it is more convenient to use the approximations in \cite[\eqns\ (11)]{Fenton1982jul} as detailed in \emph{\suppinfo \S\ref{supp:sec:approx}}, which lead to
\begin{subequations}
	\label{main:eqn:moduli:H0}
	\begin{equation}
		\label{main:eqn:krho:H0}
		\bigg. k_{\rho}^{2} \bigg|_{
			\scriptsize \shortstack{
				$H \approx 0$ \\
				$g \approx 0$
			}
		}
		\approx
		4 \coth\del{ \frac{\pi}{4} \frac{W_{AB}}{H} - \ln\sqrt{2} }
		\tanh\del{ \frac{\pi}{4} \frac{g}{H} }
	\end{equation}%
	\begin{equation}
		\label{main:eqn:krho':H0}
		\bigg. {k_{\rho}'}^{2} \bigg|_{
			\scriptsize \shortstack{
				$H \approx 0$\\
				$w_{A} \approx 0$ \\
				$w_{B} \approx 0$
			}
		}
		\approx
		16 \e^{-\pi W_{AB}/H}
		\sinh^{2}\del{ \frac{\pi}{4} \frac{w_{A}}{H} }
		\sinh^{2}\del{ \frac{\pi}{4} \frac{w_{B}}{H} }
	\end{equation}%
\end{subequations}%
See also \tab\ \ref{main:tab:K'K:approx} at \emph{Discussion \S\ref{main:sec:approximations_validity}} to know the dimensions for which these approximations hold.

\section{Discussion}

\subsection{Comparison with previous results}


The results for conformal transformation, potential distribution, flux density and flux agree with experimentally validated expressions \cite{Aoki1988dec,Igreja2004may,Morf2006may} and simulations \cite{Strutwolf2005feb,Morf2006may,GuajardoYevenes2013sep} found in the literature.
These expressions and simulations correspond to the cases of tall domain (with arbitrary band widths) and shallow domain (with equal band widths), which are particular cases of the results obtained in this report.

In case of tall domain (large $H$), the conformal transformation (\emph{Methods \S\ref{main:sec:T}}) agrees with that in \cite[\eqns\ (9), (10), (14), (15) and (19)]{Aoki1988dec}.
The potential distribution in \eqn\ \eqref{main:eqn:potential} corresponds to that in \cite[\eqns\ (20) and (21)]{Aoki1988dec}\footnote{%
	Note that \cite[$H$ in \eqn\ (21)]{Aoki1988dec} has a typo, and the $+$ sign, at the middle of the expression, should be replaced by a $-$ sign.
	For facilitating the comparisons \cite[$c^{*}H$ in \eqns\ (6) and (21)]{Aoki1988dec} and \cite[$nFD$ in \eqn\ (26)]{Aoki1988dec} correspond to $u_{B} - u_{A}$ and $\gamma$ respectively.
	The parameter \cite[$p$ in \eqn\ (15)]{Aoki1988dec} corresponds to $k_{\rho}^{2}$, which is used instead of the modulus $k_{\rho}$ as the argument for all elliptic functions.
}, which depends directly on the real part of the conformal transformation.
The flux density in \eqns\ \eqref{main:eqn:flux_density} agrees with that in \cite[\eqns\ (19) and (26)]{Aoki1988dec}, which depends directly on the imaginary part of the conformal transformation's derivative.
This derivative also agrees with that in \cite[\eqns\ (17) and (27)]{Aoki1988dec} by considering the identity $\cos(\alpha + \beta) + \cos(\alpha - \beta) = (1 + \cos(2\alpha))^{1/2} (1 + \cos(2\beta))^{1/2}$.
The flux per band in \eqns\ \eqref{main:eqn:flux:band} and their moduli in \eqns\ \eqref{main:eqn:moduli:Hinf} correspond to that in \cite[\eqns\ (15) and (28)]{Aoki1988dec} and \cite[\eqns\ (2), (3) and (6)]{Morf2006may}.
All these results agree asymptotically when $H \to +\infty$, because $k_{r} \to 0^{+}$, $K(k_{r}) \to \pi/2$ and $\cd(\cdot, k_{r}) \to \cos(\cdot)$.


\begin{figure}
	\centering
	\includegraphics{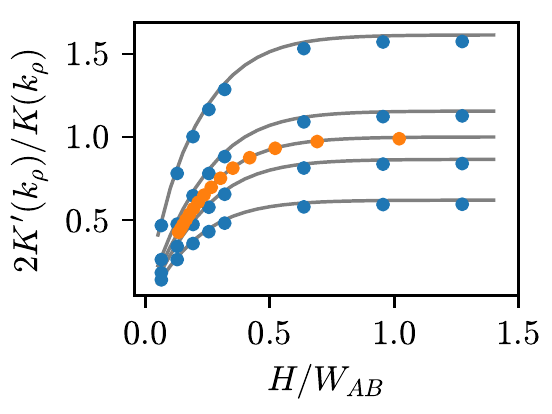}%
	\includegraphics{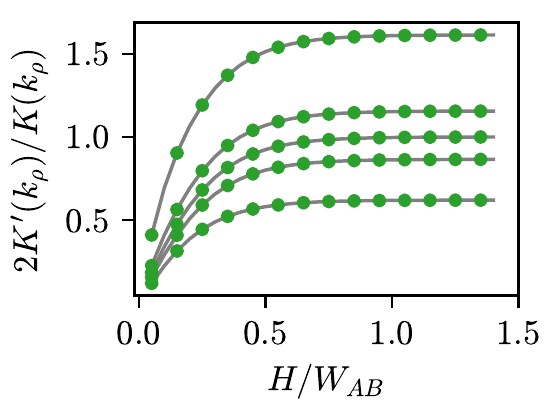}
	\caption{%
		Normalized flux per band $2 K'(k_{\rho})/K(k_{\rho}) = (i_{A}/L)/(\gamma [u_{A} - u_{B}])$ for an \emph{interior cell} with equal band widths $w_{A}/W_{AB} = w_{B}/W_{AB} \in \{ \numlist[list-final-separator={,}]{0,2; 0,4; 0,5; 0,6; 0,8} \}$ (widths correspond to solid lines in order from bottom to top).
		Solid lines: Theoretical expresion in \eqn\ \eqref{main:eqn:flux:band:int}.
		Blue \cite[\fig\ 7a]{GuajardoYevenes2013sep} and
		orange \cite[\fig\ 7a]{Strutwolf2005feb} dots: Literature simulations.
		Green dots: Theoretical expression \cite[\eqn\ (13)]{Igreja2004may}.
		See \emph{\suppinfo\ \S\ref{supp:sec:comparison}} for details on the data from the literature.
	}
	\label{main:fig:flux:comparison}
\end{figure}

In case of shallow domain, the flux per band in \eqns\ \eqref{main:eqn:flux:band} was evaluated and compared with its simulated counterparts in \cite[\fig\ 7a]{Strutwolf2005feb} \cite[\fig\ 7a]{GuajardoYevenes2013sep} and with the experimentally validated expression in \cite[\eqn\ (13)]{Igreja2004may}.
All these results agree as shown in the comparisons made in \fig\ \ref{main:fig:flux:comparison}.

\subsection{Influence of the geometry}

The conformal transformation $T_{r}^{\rho}$ in \emph{Methods \S\ref{main:sec:T}} depends only on relative dimensions and positions in the \emph{generic cell} instead of absolute ones.
The same occurs with its complex derivative in \eqn\ \eqref{main:eqn:dTdr} when this is normalized as $W \partial\bm{\rho}/\partial\bm{r}$.
This signifies that the potential $u$, normalized flux density $Wj$ and normalized flux $i_{A}/L$ in \emph{Results \S\ref{main:sec:solution}} will depend on relative dimensions of the IDA domain, and therefore, they will be shared among all IDA domains with proportional geometries.
This is because these three quantities depend directly on the real/imaginary part of either $T_{r}^{\rho}$ or $W \partial\bm{\rho} / \partial\bm{r}$.

As general behavior, the flux in \emph{interior} and \emph{exterior cells} of an IDA domain increases with both: the relative width of the bands ($w_{A}/W_{AB}$ and $w_{B}/W_{AB}$) and the relative height of the domain ($H/W_{AB}$) (\figs\ \ref{main:fig:flux} and \ref{main:fig:flux:comparison}).
In case of an \emph{exterior cell}, the flux also increases with its relative width $W/W_{AB}$, however any change appears to be comparatively small for $W/W_{AB} \gtrsim 2$.
For a fixed domain height, the isolines of flux are symmetric with respect to the main diagonal of the plot (in case of an \emph{interior cell}) and tend to collapse towards the anti-diagonal of the plot as the domain height decreases (in case of \emph{interior} and \emph{exterior cells}).

In particular, for very shallow domains ($0 < H/W_{AB} \lesssim \num{0,3}$), the flux increases slowly as the bands width increases (\figs\ \ref{main:fig:flux} and \ref{main:fig:flux:comparison}).
This is because the flux density near the center of each band is low (\fig\ \ref{main:fig:flux_density}), and therefore it cannot contribute with enough area for the flux, unless the bands are very wide.
This low flux density near the center of the bands has its origin in the potential distribution (\fig\ \ref{main:fig:potential}), which presents almost no vertical gradient due to truncation by the roof of the domain.
Most of the flux in this case comes from the flux density spikes at the edges of each band (\fig\ \ref{main:fig:flux_density}), which originate mainly due to the horizontal gradient of potential generated between neighboring bands (\fig\ \ref{main:fig:potential}).
On the other hand, the flux for very shallow domains increases quickly as the domain height increases (\figs\ \ref{main:fig:flux} and \ref{main:fig:flux:comparison}).
This is because the flux density near the center of each band increases rapidly (\fig\ \ref{main:fig:flux_density}), boosted by a quick development of the vertical gradient of potential, as the domain height increases (\fig\ \ref{main:fig:potential}).

For shallow domains ($\num{0,3} \lesssim H/W_{AB} \lesssim 1$), the flux increases more quickly as the band widths increases (\figs\ \ref{main:fig:flux} and \ref{main:fig:flux:comparison}).
This is because the flux density near the center of each band is higher (\fig\ \ref{main:fig:flux_density}), due to the vertical gradient of potential (\fig\ \ref{main:fig:potential}) that has been developed, and therefore it can contribute more area for the flux.
On the other hand, the flux for shallow domains increases more slowly as the domain height increases (\figs\ \ref{main:fig:flux} and \ref{main:fig:flux:comparison}).
The reason for this behavior is that the flux density near the center of each band also increases more slowly with the domain height (\fig\ \ref{main:fig:flux_density}), since an important part of the vertical gradient of potential has been already developed (\fig\ \ref{main:fig:potential}).

For tall domains ($H/W_{AB} \gtrsim 1$), the flux (\figs\ \ref{main:fig:flux} and \ref{main:fig:flux:comparison}) and flux density (\fig\ \ref{main:fig:flux_density}) approach their maximum limit and become independent of the domain height (they depend only on the bands width).
This is because the domain is so tall that the potential distribution becomes uniform far from the surface of the IDA, and a fully developed potential gradient is present in the domain (\fig\ \ref{main:fig:potential}).
In this case, the flux between neighboring bands is not affected by the roof of the domain and approaches that of semi-infinite geometries.

Similar results, in terms of geometry dependence, have also been found previously through simulations for planar IDA configurations \cite{Strutwolf2005feb,GuajardoYevenes2013sep,Heo2013,Heo2014jun,Kanno2014} as well as for elevated IDA configurations \cite{Goluch2009may,Heo2013,Heo2014jun}.

\subsection{Total flux at the IDA}
\label{main:sec:total_flux}

\begin{figure}
	\subcaptionbox{%
		Min: \num{0.000}.
		Max: \num{0.279}.
	}{%
		\includegraphics{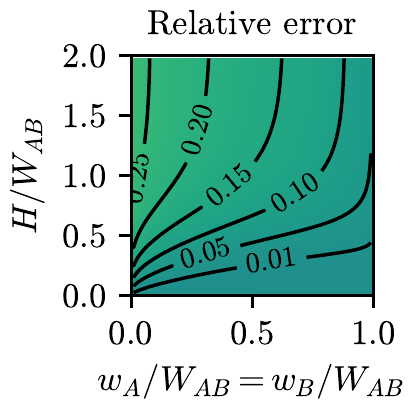}
	}%
	\subcaptionbox{%
		Min: \num{0.000}.
		Max: \num{0.522}.
	}{%
		\includegraphics{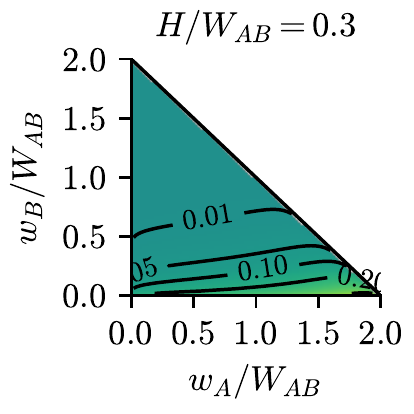}
	}%
	\subcaptionbox{%
		Min: \num{0.000}.
		Max: \num{0.619}.
	}{%
		\includegraphics{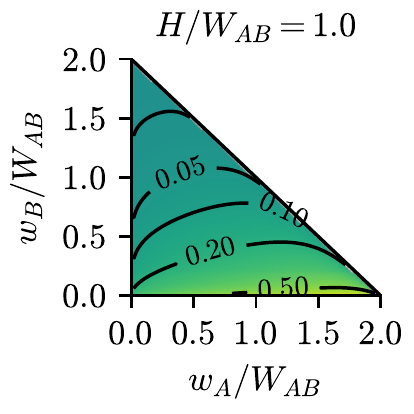}
	}%
	\caption{%
		Selected two-dimensional slices of the relative error $i_{A}^\text{ext}/i_{A}^\text{int} - 1$ as a function of the cells dimensions.
		Case of \emph{exterior cell} with $W/W_{AB} = 2$.
		Plots considering \emph{exterior cells} with $2 \leq W/W_{AB} \leq 5$ presented almost no variations (shown at \emph{\suppinfo\ \S\ref{supp:sec:approx}}).
	}
	\label{main:fig:flux:cells_error}
\end{figure}

The total flux passing through an IDA is an important quantity to consider, since in many cases it is possible to measure it experimentally.

In case of an \emph{ideal domain}, we have an assembly consisting only of \emph{interior cells} with $N_{A} = N_{B}$, thus the total flux through each array is given by
\begin{subequations}
	\label{main:eqn:flux:total}
	\begin{equation}
	-N_{A} i_{A}^\text{int} = N_{B} i_{B}^\text{int}
	\end{equation}
	
	For a \emph{practical domain}, we have an assembly of \emph{interior cells} and two \emph{exterior cells} such that $N_{A} + 1 = N_{B}$, thus the total flux through each array is given by
	\begin{equation}
		-I_{A} = I_{B} \approxeq -(N_{A} - 1) i_{A}^\text{int} - i_{A}^\text{ext}
		= (N_{B} - 2) i_{B}^\text{int} + 2 i_{B}^\text{ext}
	\end{equation}%
	If we compare this total flux with the one of an \emph{ideal domain} (both with the same number of bands for the array $A$)
	\begin{equation}
		\frac{I_{A}}{N_{A} i_{A}^\text{int}} - 1
		\approxeq \frac{1}{N_{A}} \del{ \frac{i_{A}^\text{ext}}{i_{A}^\text{int}} - 1}
	\end{equation}%
	we can see that the relative error of $I_{A}$ with respect to its \emph{ideal domain} counterpart $N_{A} i_{A}^\text{int}$ decreases towards zero as the number of bands $N_{A}$ increases.
	This shows that the flux through an \emph{interior cell} is dominant compared to that of an \emph{exterior cell}, especially as the number of bands per array increases.
	This suggests that it is possible to replace the more complex total flux of a \emph{practical domain} by that of an \emph{ideal domain}.
\end{subequations}%

In fact, for most practical cases, the relative error between the flux at \emph{exterior} and \emph{interior cells} is $|i_{A}^\text{ext}/i_{A}^\text{int} - 1| \leq \num{0,5}$ (\fig\ \ref{main:fig:flux:cells_error}),
and therefore the relative error between the total flux at \emph{practical} and \emph{ideal domains} is $|I_{A}/N_{A} i_{A}^\text{int} - 1| \lesssim \num{0,5}/N_{A}$.
This means that the number of digits in agreement for the total flux of both domains relates to the order of magnitude of the number of bands per array.
Moreover, for the cases where $2 H/W_{AB} \leq w_{A}/W_{AB} = w_{B}/W_{AB}$, we have $|i_{A}^\text{ext}/i_{A}^\text{int} - 1| < \num{0,05}$, and then it is possible to obtain an additional digit in agreement for the flux of both domains.

Considering the previous argument, we present some examples of total flux for the case of an \emph{ideal domain}. See \eqn\ \eqref{main:eqn:flux:band:int}.%
\begin{subequations}%
	In case of an IDA of \emph{electric resistance} $R$, an \emph{electric current} $I_{A}$ is produced when applying a difference of \emph{electric potentials} $V_{A} - V_{B}$ to a medium of \emph{electric conductivity}~$\sigma$
	\begin{equation}
		\frac{1}{R} = \frac{I_{A}}{V_{A} - V_{B}}
		= N_{A} L\; 2\frac{K'(k_{\rho})}{K(k_{\rho})} \bigg|_\text{int} \, \sigma
	\end{equation}%
	In case of an IDA of capacitance $C$, an \emph{electric charge} $Q_{A}$ (total flux of electric displacement) is accumulated when applying a difference of \emph{electric potentials} $V_{A} - V_{B}$ to a medium of \emph{absolute permittivity}~$\epsilon$
	\begin{equation}
		C = \frac{Q_{A}}{V_{A} - V_{B}}
		= N_{A} L\; 2\frac{K'(k_{\rho})}{K(k_{\rho})} \bigg|_\text{int} \, \epsilon
	\end{equation}%
	Note that if the IDA presents resitive and capacitive effects simultaneouly, then we have $RC = \epsilon/\sigma$
	(this relation is not limited to an IDA, but holds in general \cite{Olthuis1995mar}).
	In case of electrochemistry, a \emph{faradaic current} $I_{A}$ is produced when there is a difference of \emph{concentrations} $c_{A} - c_{B}$ of electrochemical species at the bands of the IDA
	\begin{equation}
		I_{A}
		= N_{A} L\; 2\frac{K'(k_{\rho})}{K(k_{\rho})} \bigg|_\text{int} \, F n_{e} D \, [c_{A} - c_{B}]
	\end{equation}%
	Here $F$ corresponds to the Faraday's constant, and the electrochemical species exchange $n_{e}$ electrons in the faradaic process and have a \emph{diffusion coefficient} $D$ in the medium.
\end{subequations}

\subsection{IDA of minimal surface}

The amount of material used for fabricating an IDA may be of importance in processes of additive manufacturing.
This amount is proportional to $m = w_{A}/W_{AB} + w_{B}/W_{AB}$ and can be minimized with ease in case of an \emph{ideal domain} or a \emph{practical domain} with a large amount of bands (\fig\ \ref{main:fig:flux:interior}).

For fixed domain height and a desired flux, the minimum is obtained when $w_{A} = w_{B}$.
This can be seen by selecting a desired isoline of flux, and intersecting it with the anti-diagonal $m = w_{A}/W_{AB} + w_{B}/W_{AB}$.
This produces two intersection points, which are symmetric with respect to the diagonal line  $w_{A}/W_{AB} = w_{B}/W_{AB}$.
The minimum is obtained by moving the anti-diagonal towards the origin, which makes the two intersection points on the isoline converge to a single point, thus obtaining $w_{A} = w_{B}$.

\subsection{Regions of validity for flux approximations}
\label{main:sec:approximations_validity}

\begin{figure}[t]
	\centering
	\includegraphics{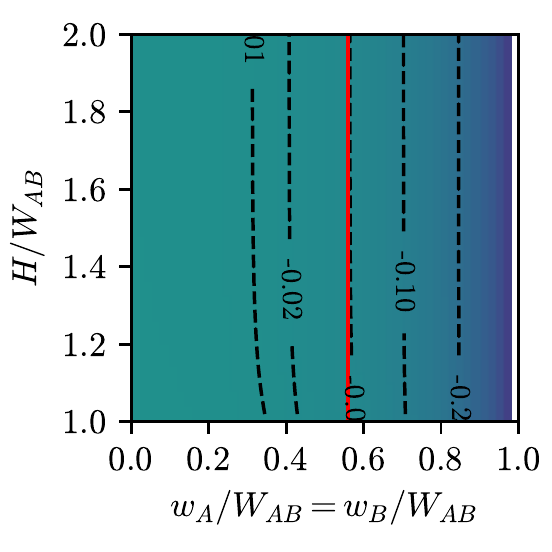}
	\includegraphics{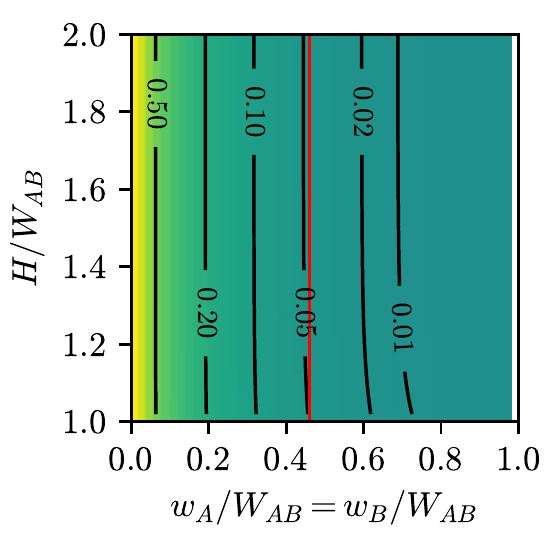}
	\\
	\includegraphics{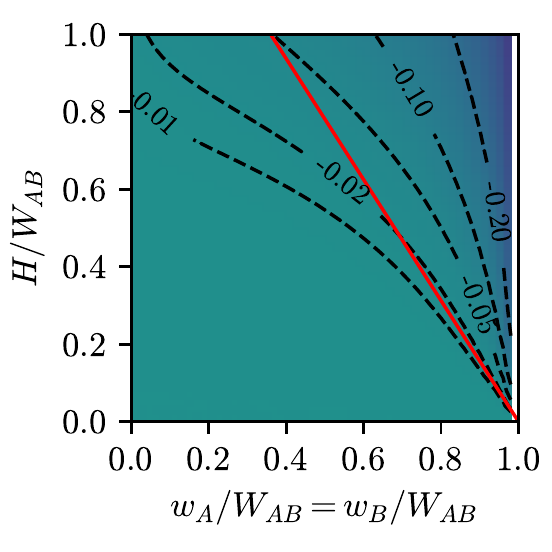}
	\includegraphics{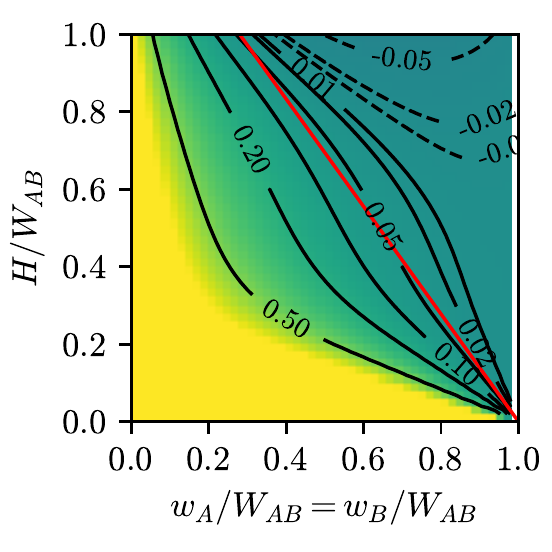}
	\caption{%
		Relative error of the approximated ratio $K'(k_{\rho})/K(k_{\rho})$ for an \emph{ideal domain} with respect to its exact counterpart.
		Top and bottom: Errors for tall and shallow domains respectively.
		Left and right: Errors for narrow and wide bands respectively.
		In red: straight lines close to errors of \SI{+-5}{\percent}.
	}
\end{figure}

The search for the regions of validity for \eqns\ \eqref{main:eqn:K'K:approx} \eqref{main:eqn:moduli:Hinf} and \eqref{main:eqn:moduli:H0} was done graphically, and therefore, it was more convenient to consider bands of equal width, such that the resulting plots were two-dimensional.

Approximations of $K'(k_{\rho})/K(k_{\rho})$ in \eqns\ \eqref{main:eqn:K'K:approx} and \eqref{main:eqn:moduli:Hinf} for tall domains ($H/W_{AB} > 1$) were obtained first by Aoki and colleagues, for the case of wide bands ($g \approx 0$) \cite[Eq. (32)]{Aoki1988dec}, and later by Morf and colleagues, for the case of narrow bands \cite[Eqs. (2), (6) and (7)]{Morf2006may}.
The approximation for wide bands becomes accurate (error less than \SI{+-5}{\percent}) for bands of relative width $> \num{0,46}$.
In the case of narrow bands, the approximation has an error less than \SI{+-5}{\percent} for bands of relative width $< \num{0,56}$.
The regions of approximation for both sizes of bands overlap, thus covering all cases of $H/W_{AB} > 1$ with a relative error less than \SI{+-5}{\percent}.
See \tab\ \ref{main:tab:K'K:approx} for a summary.

Approximations of $K'(k_{\rho})/K(k_{\rho})$ in Eqs. \eqref{main:eqn:K'K:approx} and \eqref{main:eqn:moduli:H0} for shallow domains ($H/W_{AB} \leq 1$) are results that have not been published before.
The approximation for wide bands becomes accurate (error approximately less than \SI{+-5}{\percent}) for bands of relative width approximately larger than $1 - \num{0,72}H/W_{AB}$.
In the case of narrow bands, the approximation has an error less than \SI{+-5}{\percent} for bands of relative width approximately smaller than $1 - \num{0,64}H/W_{AB}$.
The regions of approximation for both sizes of electrodes overlap, thus covering all cases of $H/W_{AB} \leq 1$ with a relative error less than \SI{+-5}{\percent}.
See Table \ref{main:tab:K'K:approx} for a summary.

\begin{table}
	\centering
	\begin{tabular}{cc}
	\toprule
	Domain & Approximation \\
	\midrule
	\multicolumn{1}{l}{\emph{Case of tall domains} ($H/W_{AB} > 1$)} \\
	$w_{E}/W_{AB} > \num{0,46}$ &
	\eqns\ \eqref{main:eqn:K'K:k0}, \eqref{main:eqn:krho:Hinf} \\
	$w_{E}/W_{AB} < \num{0,56}$ &
	\eqns\ \eqref{main:eqn:K'K:k'0}, \eqref{main:eqn:krho':Hinf} \\
	\midrule
	\multicolumn{1}{l}{\emph{Case of shallow domains} ($H/W_{AB} \leq 1$)} \\
	$w_{E}/W_{AB} + \num{0,72}H/W_{AB} \gtrsim 1$ &
	\eqns\ \eqref{main:eqn:K'K:k0}, \eqref{main:eqn:krho:H0} \\
	$w_{E}/W_{AB} + \num{0,64}H/W_{AB} \lesssim 1$ &
	\eqns\ \eqref{main:eqn:K'K:k'0}, \eqref{main:eqn:krho':H0} \\
	\bottomrule
\end{tabular}%

	\caption{%
		Regions where the approximations hold with a relative error less than \SI{+-5}{\percent}.
		Here $w_{E} := w_{A} = w_{B}$.
		See \emph{\suppinfo\ \S\ref{supp:sec:regions}} for details.
	}
	\label{main:tab:K'K:approx}
\end{table}

\section{Conclusion}



It is possible to transform \emph{ideal} and \emph{practical} domains containing an IDA into a parallel-plates domain.
In the latter, the solution of Laplace's equation is simple, and corresponds to a linear interpolation of the potentials applied at both plates.
This solution was transformed back into the IDA domain, leading to an analytical  expression for the potential distribution (based on elliptic functions and integrals) from which flux density and flux were derived.


All three expressions depend on relative dimensions of the IDA domain instead of absolute ones, meaning that the same results hold for proportionally similar IDAs.
The flux density and flux at the IDA increase as the bands width and domain height increase, which correlates with an increase of gradients in the potential distribution.
Their behavior approaches that of an IDA in a semi-infinite domain, once the domain is sufficiently tall
(approximately when the domain height is greater than the separation between centers of consecutive bands).


Among the three quantities, the flux plays an important role, since in many cases it is accessible for experimental measurement.
Therefore, its analytical expression can be used during the design process of an IDA or as a benchmark of performance in measurements.
In case of an \emph{ideal domain}, its analytical expression holds exactly, while that of a \emph{practical domain} approaches the latter as the number of bands per array increases
(the number of digits in agreement between both expressions correlates  with the order of magnitude of the number of bands per array).
This is because the flux in \emph{interior cells} of an IDA becomes dominant over that in \emph{exterior cells}, making fringing effects negligible.


For an IDA behaving like one in an \emph{ideal domain}, we found that bands of equal width minimize its total surface for any desired flux.
Also, approximations for the exact flux were found.
Elementary functions were used to approximate the cases of tall and shallow domains respectively, producing expressions which are accurate with a relative error smaller than \SI{+-5}{\percent} with respect to their exact counterpart.
When these approximations are used in combination, they cover all possibilities of interest for IDAs in confined domains.

\section{Acknowledgements}

The authors deeply appreciate the aid and comments of
Mithran Somasundrum and Sirimarn Ngamchana,
which helped to improve the quality of this manuscript.
\textsc{cfgy} gratefully acknowledges
\emph{Petchra Pra Jom Klao Ph.~D. scholarship} (Grant No. 28/2558),
\emph{King Mongkut’s University of Technology Thonburi}.
The authors acknowledge the financial support provided by
\emph{King Mongkut’s University of Technology Thonburi}
through the \emph{KMUTT 55th Anniversary Commemorative Fund},
%
and King Mongkut’s University and the \emph{Research Network NANOTEC} (RNN) program (grant no. P1851883)
of the \emph{National Nanotechnology Center} (NANOTEC), NSTDA,
Ministry of Science and Technology, Thailand.



\section{\suppinfo}

(\textsc{pdf}) Details on calculations, additional figures and tables.
(\textsc{zip}) \textsf{Python} scripts for numerical calculations and plots \cite{GuajardoYevenes2021elektrodo}.

\renewcommand{\refname}{\vspace*{-1.5em}}
\section{References}
\bibliographystyle{utphys-cfgy}
\bibliography{refs-main-urls}

\startsuppinfo
\author{
	\authorigivennames\ \textsc{\authorifamilynames}
	\orcid{\authoriorcid}\arxiv{\authoriarxiv}
	\\ \texttt{\authoriemail}
	\and
	\authoriigivennames\ \textsc{\authoriifamilynames}
	\orcid{\authoriiorcid}
	\\ \texttt{\authoriiemail}
}

\maketitle

\section[Transformation of generic cell into parallel plates]{Transformation of the generic cell of IDA domain to parallel plates}
\label{supp:sec:T}

Alternated potential and insulation boundaries at the bottom of the cells make it difficult to obtain an analytical solution for Laplace's equation.
Nevertheless, it is possible to arrange these alternated boundary conditions, so they can be placed at different walls in a transformed cell, by using conformal transformations.

Complex conformal transformations leave Laplace's equation, as well as potential (Dirichlet) and non-flux (Neumann) boundary conditions, invariant under domain changes \cite[\S5.7]{Driscoll2002}.
In particular, Jacobian elliptic functions $\sn()$ and $\cd()$ are of interest, since they conformally transform a square domain into the upper half-plane \cite[\S2.5]{Driscoll2002} \cite[\S\dlmf{22.18.}{ii}]{dlmf}.
Möbius functions are also important, since they reorganize the upper half-plane,
by mapping it into itself \cite[\S2.3]{Driscoll2002}.

\begin{figure*}
	\centering
	\includegraphics{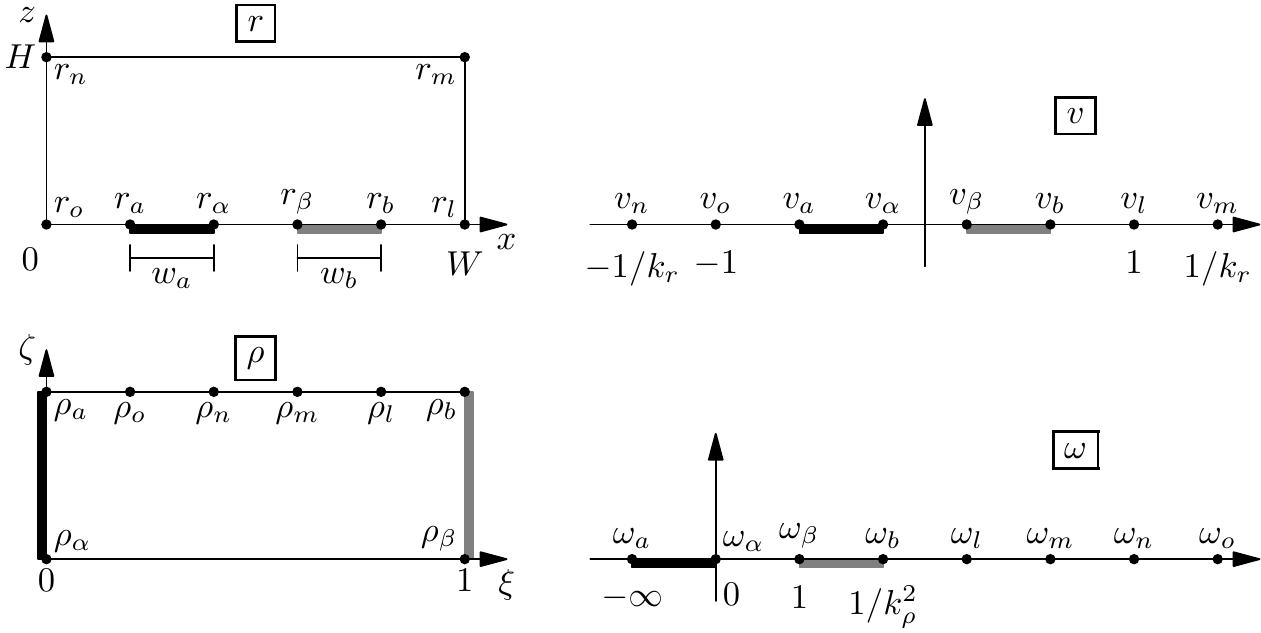}
	\caption{%
		Complex transformation $\bm{\rho} = T_{r}^{\rho}(\bm{r})$
		from the generalized IDA domain $\bm{r}=(x,z)$
		into the conformal parallel-plates domain $\bm{\rho}=(\xi,\zeta)$,
		by using the auxiliary complex domains $\bm{v}$ and $\bm{\omega}$.
	}
	\label{supp:fig:T}
\end{figure*}

Now we take a \emph{generic cell} (\fig\ \ref{supp:fig:T}) representing simultaneously \emph{interior} and \emph{exterior cells} of an IDA, depending on how its parameters are chosen.
In case of an \emph{interior cell}, one chooses
\begin{subequations}
	\label{supp:eqn:cell:int-ext}
	\begin{equation}
		\bm{r}_{o} = \bm{r}_{a} = 0
		,\quad w_{a} = w_{A}/2
		,\quad w_{b} = w_{B}/2
		,\quad \bm{r}_{b} = \bm{r}_{l} = W_{AB}
	\end{equation}%
	Whereas, in case of an \emph{exterior cell}
	\begin{equation}
		\bm{r}_{o} = \bm{r}_{a} = 0
		,\quad w_{a} = w_{A}/2
		,\quad w_{b} = w_{B}
		,\quad \bm{r}_{b} = W_{AB} + w_{B}/2 < \bm{r}_{l} = W
	\end{equation}%
\end{subequations}%

Let $\bm{r} = x + \bm{i}z$ designate the coordinates for the \emph{generic cell} of the IDA domain, where $\bm{i}^{2} = -1$.
We aim to find a complex function $\bm{\rho} = T_{r}^{\rho}(\bm{r})$ that can transform the cell from the IDA domain into a parallel-plates domain of coordinates $\bm{\rho} = \xi + \bm{i}\zeta$, as shown in \fig\ \ref{supp:fig:T}.

The transformation $T_{r}^{\rho}$ will be decomposed in three stages
\begin{equation}
	T_{r}^{\rho} = T_{\omega}^{\rho} \circ T_{v}^{\omega} \circ T_{r}^{v}
\end{equation}
First, $T_{r}^{v}$ will transform the IDA domain $\bm{r}$ to the upper half-plane $\bm{v}$ by using the elliptic function $\cd()$.
Later, $T_{v}^{\omega}$ will reorganize the upper half-plane $\bm{v}$ into the upper half-plane $\bm{\omega}$ by using a Möbius function.
Finally, $T_{\omega}^{\rho}$ will transform the upper half-plane $\bm{\omega}$ into the parallel plates domain $\bm{\rho}$ by using a combination of square root and inverse elliptic function $\arcsn()$.

\subsection{From \texorpdfstring{$\bm{r}$}{r} to \texorpdfstring{$\bm{v}$}{v} domain}

Considering the special values of the function $\arcsn()$ \citex{\tab\ \dlmf[T]{22.5.}{1}}{dlmf}
\begin{subequations}
	\label{supp:eqn:arcsn:values}
	\begin{align}
		0 &= \arcsn(0,k) \\
		\label{supp:eqn:K}
		\pm K(k) &= \arcsn(\pm 1,k) \\
		\pm K(k) + \bm{i} K'(k) &= \arcsn(\pm 1/k,k) \\
		\bm{i} K'(k) &= \arcsn(\bm{\infty},k)
	\end{align}%
\end{subequations}%
one can construct a function $(T_{r}^{v})^{-1}$ that maps the upper half-plane of $\bm{v}$ into the IDA domain $\bm{r}$, as shown in \fig\ \ref{supp:fig:T}.
The scale and translation of the function $\arcsn()$ must be chosen such that $\bm{v}_{o} = -1$ is mapped to $\bm{r}_{o} = 0$ and $\bm{v}_{l} = 1$ is mapped to $\bm{r}_{l} = W$, which leads to
\begin{subequations}
	\begin{gather}
		\bm{r} = (T_{r}^{v})^{-1}(\bm{v}) 
		= \frac{W}{2} \sbr{
			1 + \frac{1}{K(k_{r})}\arcsn(\bm{v},k_{r})
		}
		\\
		\bm{v} = T_{r}^{v}(\bm{r}) 
		= \sn\!\del{ K(k_{r})\frac{2\bm{r}}{W} - K(k_{r}), k_{r} }
		= -\cd\!\del{ K(k_{r})\frac{2\bm{r}}{W}, k_{r} }
		\label{supp:eqn:r-v}
	\end{gather}%
\end{subequations}%
due to the quarter- and half-period properties \citex{\tab\ \dlmf[T]{22.4.}{3}}{dlmf}
\begin{equation}
	\label{supp:eqn:sn-cd}
	\sn(\bm{u} - K(k),k) = \sn(\bm{u} + 3K(k),k)
	= \sn(\bm{u} + K(k) + 2K(k),k)
	= -\cd(\bm{u},k)
\end{equation}
where $4K(k)$ corresponds to one period.

The appropriate modulus $k_{r}$ can be obtained by forcing $\bm{v}_{n} = -1/k_{r}$ to be mapped to the upper-left corner $\bm{r}_{n} = \bm{i}H$
\begin{equation*}
	\bm{r}_{n} = (T_{r}^{v})^{-1}(\bm{v}_{n})
	\Leftrightarrow
	\bm{i}H = \bm{i} \frac{W}{2} \frac{K'(k_{r})}{K(k_{r})}
\end{equation*}
This leads to a \emph{lattice parameter} $\tau_{r}$ \citex{\S\dlmf{20.}{1}}{dlmf}
\begin{equation}
	\label{supp:eqn:K1Kr}
	\tau_{r} = \bm{i} \frac{K'(k_{r})}{K(k_{r})} = \bm{i} \frac{2H}{W}
\end{equation}
or, equivalently, to the \emph{elliptic nome} and its associated counterpart
\begin{subequations}
	\label{supp:eqn:qr-qr'}
	\begin{align}
		q_{r} = Q(k_{r}) &= \exp\del{ -\pi \frac{2H}{W} } \\
		q'_{r} = Q(k'_{r}) &= \exp\del{ -\pi \frac{W}{2H} }
	\end{align}%
\end{subequations}%
by applying the \emph{nome function} $Q(k)$, which is defined by the relations \cite[\S VI.3 \eqn\ (16)]{Nehari1952} \citex{\eqns\ (\dlmf[E]{19.2.}{9}) and (\dlmf[E]{22.2.}{1})}{dlmf}
\begin{equation}
	\label{supp:eqn:nomo}
	\frac{\ln Q(k)}{-\pi} = \frac{-\pi}{\ln Q(k')} = \frac{K'(k)}{K(k)}
\end{equation}

The points $\bm{v}_{o} = -1$, $\bm{v}_{l} = 1$ and $\bm{v}_{n} = -1/k_{r}$,
define completely the function $T_{r}^{v}$ in \eqn\ \eqref{supp:eqn:r-v},
which transforms the IDA domain $\bm{r}$ into the upper half-plane $\bm{v}$
as shown in \fig\ \ref{supp:fig:T}.
Thus all points of interest $\bm{v}_{a}$, $\bm{v}_{\alpha}$, $\bm{v}_{\beta}$
and $\bm{v}_{b}$ are obtained by evaluating \eqn\ \eqref{supp:eqn:r-v}
\begin{subequations}
	\label{supp:eqn:v:AB}
	\begin{align}
		\bm{v}_{a} = -\cd\!\del{ K(k_{r})\frac{2 \bm{r}_{a}}{W}, k_{r} },
		& \quad
		\bm{v}_{\alpha}
		= -\cd\!\del{ K(k_{r})\frac{2(\bm{r}_{a} + w_{a})}{W}, k_{r} }
		\\
		\bm{v}_{b} = -\cd\!\del{ K(k_{r})\frac{2 \bm{r}_{b}}{W}, k_{r} },
		& \quad
		\bm{v}_{\beta}
		= -\cd\!\del{ K(k_{r})\, \frac{2(\bm{r}_{b} - w_{b})}{W}, k_{r} }
	\end{align}	
\end{subequations}

\subsection{From \texorpdfstring{$\bm{v}$}{v} to \texorpdfstring{$\bm{\omega}$}{ω} domain}


One can use a Möbius transformation $T_{v}^{\omega}$ for reorganizing the upper half-plane of $\bm{v}$ into the upper half-plane of $\bm{\omega}$.
See \fig\ \ref{supp:fig:T}.
This transformation is constructed such that:
(i) it maps the band interval $\bm{v} \in [\bm{v}_{a}, \bm{v}_{\alpha}]$ to the negative real axis of $\bm{\omega}$ and the band interval $\bm{v} \in [\bm{v}_{\beta}, \bm{v}_{b}]$ to the real interval $\bm{\omega} \in [1,\bm{\omega}_{b}]$, and
(ii) it ensures the mapping of the upper half-plane of $\bm{v}$ into the upper half-plane of $\bm{\omega}$.

\begin{subequations}
	Condition (i) can be satisfied by choosing the Möbius function as below
	\begin{equation}
		\label{supp:eqn:v-omega}
		\bm{\omega} = T_{v}^{\omega}(\bm{v}) = 
		\frac{(\bm{v} - \bm{v}_{\alpha})}{(\bm{v} - \bm{v}_{a})}
		\frac{(\bm{v}_{\beta} - \bm{v}_{a})}{(\bm{v}_{\beta} - \bm{v}_{\alpha})}
	\end{equation}%
	since this maps $\bm{v}_{a}$, $\bm{v}_{\alpha}$ and $\bm{v}_{\beta}$ into $\bm{\omega}_{a} = \bm{\infty}$, $\bm{\omega}_{\alpha} = 0$ and $\bm{\omega}_{\beta} = 1$.
	Condition (ii) can be ensured by rewritting $T_{v}^{\omega}$ as a composition of translations, rotations/scalings, and inversion
	\begin{equation}
		\bm{\omega} = T_{v}^{\omega}(\bm{v})
		= \left[
		1 + \frac{\bm{v}_{a} - \bm{v}_{\alpha}}{\bm{v} - \bm{v}_{a}}
		\right]
		\underbrace{\left[
			\frac{\bm{v}_{\beta}-\bm{v}_{a}}{\bm{v}_{\beta}-\bm{v}_{\alpha}}
			\right]}_{p}
	\end{equation}%
\end{subequations}%
The fact that $\bm{v}_{a} - \bm{v}_{\alpha} < 0$ and $p > 0$ ensures that the upper half-plane of $\bm{v}$ is mapped to the upper half-plane of $\bm{\omega}$.
See \cite[\eqn\ (5.7.3)]{Ablowitz2003apr}, \cite[\S V.2 \eqns\ (6), (7) and (10)]{Nehari1952} or \cite[Examples 5.3, 5.4 and 5.7]{Olver2020jun} for more details on decomposition and mapping of Möbius functions.

\begin{subequations}
	Here most points of interest are already known from the definition of $T_{v}^{\omega}$
	\begin{equation}
		\bm{\omega}_{a} = \infty, \quad
		\bm{\omega}_{\alpha} = 0, \quad
		\bm{\omega}_{\beta} = 1, \quad
	\end{equation}%
	and the remaining point $\bm{\omega}_{b}$ can be obtained by direct evaluation
	\begin{equation}
		\bm{\omega}_{b} =
		\frac{(\bm{v}_{b} - \bm{v}_{\alpha})}{(\bm{v}_{b} - \bm{v}_{a})}
		\frac{(\bm{v}_{\beta} - \bm{v}_{a})}{(\bm{v}_{\beta} - \bm{v}_{\alpha})}
	\end{equation}%
\end{subequations}%

\subsection{From \texorpdfstring{$\bm{\omega}$}{ω} to \texorpdfstring{$\bm{\rho}$}{ρ} domain}

The last transformation $T_{\omega}^{\rho}$ is in charge of mapping the upper half-plane of $\bm{\omega}$ into the parallel-plates domain $\bm{\rho}$.
See \fig\ \ref{supp:fig:T}.
This is achieved in two stages:
(i) The upper half-plane of $\bm{\omega}$ is mapped to the first quadrant of $\sqrt{\bm{\omega}}$,
such that the negative real axis of $\bm{\omega}$ is mapped to the positive imaginary axis of $\sqrt{\bm{\omega}}$
and the real interval $\bm{\omega} \in [1,\bm{\omega}_{b}]$ is mapped to the real interval $\sqrt{\bm{\omega}} \in [1,\sqrt{\bm{\omega}_{b}}]$.
(ii) The first quadrant of $\sqrt{\bm{\omega}}$ is mapped to the parallel-plates domain $\bm{\rho}$ by using the special values of $\arcsn()$ in Eqs. \eqref{supp:eqn:arcsn:values} or \citex{\tab\ \dlmf[T]{22.5.}{1}}{dlmf}.

The scaling and translation of the function $\arcsn()$ are chosen such that $\sqrt{\bm{\omega}_{\alpha}}=0$ is mapped to $\bm{\rho}_{\alpha}=0$ and $\sqrt{\bm{\omega}_{\beta}}=1$ is mapped to $\bm{\rho}_{\beta}=1$, which leads to
\begin{equation}
	\label{supp:eqn:omega-rho}
	\bm{\rho} = T_{\omega}^{\rho}(\bm{\omega})
	= \frac{1}{K(k_{\rho})} \arcsn(\sqrt{\bm{\omega}}, k_{\rho})
\end{equation}
The appropriate modulus $k_{\rho}$ is obtained by choosing $\sqrt{\bm{\omega}_{b}} = 1/k_{\rho}$, such that $\sqrt{\bm{\omega}_{b}}$ be mapped to the upper right corner $\bm{\rho}_{b}$ of the parallel-plates domain, which leads to the following expressions for the modulus and complementary modulus
\begin{subequations}
	\label{supp:eqn:krho-krho'}
	\begin{align}
		k_{\rho}^{2} = \frac{1}{\bm{\omega}_{b}} &=
		\frac{(\bm{v}_{b} - \bm{v}_{a})}{(\bm{v}_{b} - \bm{v}_{\alpha})}
		\frac{(\bm{v}_{\beta} - \bm{v}_{\alpha})}{(\bm{v}_{\beta} - \bm{v}_{a})}
		\\
		{k_{\rho}'}^{2} = 1 - k_{\rho}^{2} &=
		\frac{(\bm{v}_{b} - \bm{v}_{\beta})}{(\bm{v}_{b} - \bm{v}_{\alpha})}
		\frac{(\bm{v}_{\alpha} - \bm{v}_{a})}{(\bm{v}_{\beta} - \bm{v}_{a})}
	\end{align}%
\end{subequations}%

Two points of interest, $\bm{\rho}_{\alpha}$ and $\bm{\rho}_{\beta}$, are already known from the definition of $T_{\omega}^{\rho}$.
The other two points, $\bm{\rho}_{a}$ and $\bm{\rho}_{b}$, are defined when choosing the modulus $k_{\rho}$ and by using the special values of $\arcsn()$ in \eqns\ \eqref{supp:eqn:arcsn:values} or \citex{\tab\ \dlmf[T]{22.5.}{1}}{dlmf}
\begin{subequations}
	\begin{align}
		\bm{\rho}_{a} = \xi_{a} + \bm{i} \zeta_{a}
		= \frac{1}{K(k_{\rho})} \arcsn(\sqrt{\bm{\omega}_{a}}, k_{\rho})
		= 0 + \bm{i} \frac{K'(k_{\rho})}{K(k_{\rho})},
		& \quad
		\bm{\rho}_{\alpha} = 0
		\\
		\bm{\rho}_{b} = \xi_{b} + \bm{i} \zeta_{b}
		= \frac{1}{K(k_{\rho})} \arcsn(\sqrt{\bm{\omega}_{b}}, k_{\rho})
		= 1 + \bm{i} \frac{K'(k_{\rho})}{K(k_{\rho})},
		& \quad
		\bm{\rho}_{\beta} = 1
	\end{align}%
\end{subequations}%

\subsection{Derivative of the domain transformation}
\label{supp:sec:dTdr}

For convenience, we first summarize the domain transformation
\begin{subequations}
	\begin{align}
		\bm{\rho} &= T_{r}^{\rho}(\bm{r})
		= T_{\omega}^{\rho} \circ T_{v}^{\omega} \circ T_{r}^{v}(\bm{r})
		\\
		\bm{\rho} &= T_{\omega}^{\rho}(\bm{\omega})
		= \frac{1}{K(k_{\rho})} \arcsn(\sqrt{\bm{\omega}}, k_{\rho})
		\\
		\bm{\omega} &= T_{v}^{\omega}(\bm{v}) = 
		\frac{(\bm{v} - \bm{v}_{\alpha})}{(\bm{v} - \bm{v}_{a})}
		\frac{(\bm{v}_{\beta} - \bm{v}_{a})}{(\bm{v}_{\beta} - \bm{v}_{\alpha})}
		\\
		\bm{v} &= T_{r}^{v}(\bm{r}) 
		= -\cd\!\del{ K(k_{r})\frac{2\bm{r}}{W}, k_{r} }
	\end{align}%
\end{subequations}%

To obtain its derivative, we decompose it by taking the derivatives for each of the transformation steps
\begin{subequations}
	\begin{align}
		\dpd{\bm{\rho}}{\bm{r}} &=
		\dpd{\bm{\rho}}{\bm{\omega}}
		\dpd{\bm{\omega}}{\bm{v}}
		\dpd{\bm{v}}{\bm{r}}
		\\
		\dpd{\bm{\rho}}{\bm{\omega}} &=
		\frac{1}{2 K(k_{\rho})}\,
		\frac{1}{
			\bm{\omega}^{1/2} (1-\bm{\omega})^{1/2} (1-k_{\rho}^{2}\bm{\omega})^{1/2}
		}
		\\
		\dpd{\bm{\omega}}{\bm{v}} &=
		\frac{
			\textcolor{Blue1}{(\bm{v}_{\alpha}-\bm{v}_{a})}
		}{
			\textcolor{Red2}{(\bm{v}-\bm{v}_{a})^{2}}
		}
		\frac{
			\textcolor{Yellow4}{(\bm{v}_{\beta}-\bm{v}_{a})}
		}{
			\textcolor{Green4}{(\bm{v}_{\beta}-\bm{v}_{\alpha})}
		}
		\\
		\label{supp:eqn:dTdr:v-r}
		\dpd{\bm{v}}{\bm{r}} &=
		\frac{2 K(k_{r})}{W}
		\, (1 - \bm{v}^{2})^{1/2} (1 - k_{r}^{2} \bm{v}^{2})^{1/2}
	\end{align}%
\end{subequations}%
where $\partial \bm{\rho}/\partial \bm{\omega}$ is due to $\partial \arcsn(\bm{u}, k)/\partial \bm{u} = (1 - \bm{u}^{2})^{-1/2} (1 - k^{2} \bm{u}^{2})^{-1/2}$ in \citex{\eqn\ (\dlmf[E]{22.15.}{12})}{dlmf},
and $\partial \bm{v}/\partial \bm{r}$ is due to the combination of $\partial \cd(\bm{u}, k)/\partial \bm{u} = -{k'}^{2} \sd(\bm{u}, k) \nd(\bm{u}, k)$ in \citex{\tab\ \dlmf[T]{22.13.}{1}}{dlmf} and ${k'}^{2} \sd(\bm{u}, k)^{2} = 1 - \cd(\bm{u}, k)^{2}$ and ${k'}^{2} \nd(\bm{u}, k)^{2} = 1 - k^{2} \cd(\bm{u}, k)^{2}$ in \citex{\eqn\ (\dlmf[E]{22.6.}{4})}{dlmf}.
The value of each factor in the denominator of $\partial \bm{\rho}/\partial \bm{\omega}$ is
\begin{subequations}
	\begin{align}
		\bm{\omega} &=
		\frac{
			(\bm{v}-\bm{v}_{\alpha}) \textcolor{Yellow4}{(\bm{v}_{\beta}-\bm{v}_{a})}
		}{
			\textcolor{Red2}{(\bm{v}-\bm{v}_{a})} \textcolor{Green4}{(\bm{v}_{\beta}-\bm{v}_{\alpha})}
		}
		\\
		(1 - \bm{\omega}) &=
		-\frac{
			(\bm{v}-\bm{v}_{\beta}) \textcolor{Blue1}{(\bm{v}_{\alpha}-\bm{v}_{a})}
		}{
			\textcolor{Red2}{(\bm{v}-\bm{v}_{a})} \textcolor{Green4}{(\bm{v}_{\beta}-\bm{v}_{\alpha})}
		}
		\\
		(1 - k_{\rho}^{2}\bm{\omega}) &=
		\frac{
			(\bm{v}_{b}-\bm{v}) \textcolor{Blue1}{(\bm{v}_{\alpha}-\bm{v}_{a})}
		}{
			\textcolor{Red2}{(\bm{v}-\bm{v}_{a})} (\bm{v}_{b}-\bm{v}_{\alpha})
		}
	\end{align}%
\end{subequations}%
since $1/k_{\rho}^{2} = \bm{\omega}_{b} = T_{v}^{\omega}(\bm{v}_{b})$.

Later, we combine the first two derivatives $\partial \bm{\rho}/\partial \bm{\omega}$ and $\partial \bm{\omega}/\partial \bm{v}$
\begin{equation*}
	\label{supp:eqn:dTdr:rho-v}
	\dpd{\bm{\rho}}{\bm{\omega}}
	\dpd{\bm{\omega}}{\bm{v}}
	=
	\frac{\pm\bm{i}}{2 K(k_{\rho})}\,
	\frac{
		(\bm{v}_{b}-\bm{v}_{\alpha})^{1/2}
		\textcolor{Yellow4}{(\bm{v}_{\beta}-\bm{v}_{a})^{1/2}}
	}{
		(\bm{v}-\bm{v}_{\alpha})^{1/2} (\bm{v}-\bm{v}_{\beta})^{1/2}
	}
	\frac{1}{
		\textcolor{Red2}{(\bm{v}-\bm{v}_{a})^{1/2}}
		(\bm{v}_{b}-\bm{v})^{1/2}
	}
\end{equation*}
which leads to an expression that consists of two complex branches: $\pm \bm{i}$.

By introducing the third derivative $\partial \bm{v}/\partial \bm{r}$ we arrive to the final expression
\begin{equation}
	\label{supp:eqn:dTdr}
	\dpd{\bm{\rho}}{\bm{r}}
	=
	\frac{\pm \bm{i}}{W} \frac{K(k_{r})}{K(k_{\rho})}\,
	\frac{
		(\bm{v}_{b}-\bm{v}_{\alpha})^{1/2} 
		(\bm{v}_{\beta}-\bm{v}_{a})^{1/2}
	}{
		(\bm{v}-\bm{v}_{\alpha})^{1/2}
		(\bm{v}-\bm{v}_{\beta})^{1/2}
	}
	\frac{
		(1 - \bm{v}^{2})^{1/2}
		(1 - k_{r}^{2} \bm{v}^{2})^{1/2}
	}{
		(\bm{v}-\bm{v}_{a})^{1/2}
		(\bm{v}_{b}-\bm{v})^{1/2}
	}
\end{equation}
In case $u_{B} > u_{A}$, the flux density at each band $B$ must be positive
\begin{equation}
	j(x) = \gamma [u_{B} - u_{A}] \Im \dpd{\bm{\rho}}{\bm{r}}(x) \geq 0,
	\text{ where the physical constant } \gamma > 0
\end{equation}
In order to achieve this, the $+\bm{i}$ branch of $\partial\bm{\rho}/\partial\bm{r}$ in \eqn\ \eqref{supp:eqn:dTdr} must be chosen as the one carrying physical meaning.
This is because $1/k_{r} > 1 \geq \bm{v}_{b} > \bm{v} > \bm{v}_{\beta} > \bm{v}_{\alpha} > \bm{v}_{a}$
when $\bm{v} \in B$, as it can be seen in \fig\ \ref{supp:fig:T}.

\section{Flux per band for interior and exterior cells}
\label{supp:sec:flux}

\begin{figure}
	\subcaptionbox{%
		Min: \num{0.049}.
		Max: \num{3.726}.
	}{%
		\includegraphics{fig-flux-W20-Hall}
	}%
	\subcaptionbox{%
		Min: \num{0.325}.
		Max: \num{3.200}.
	}{%
		\includegraphics{fig-flux-W20-H03}
	}%
	\subcaptionbox{%
		Min: \num{0.407}.
		Max: \num{3.743}.
	}{%
		\includegraphics{fig-flux-W20-H10}
	}%
	\caption{%
		Normalized flux per band $2K'(k_{\rho})/K(k_{\rho}) = (i_{A}/L)/(\gamma [u_{A} - u_{B}])$ as a function of the cell dimensions.
		Case of \emph{exterior cell} with $W/W_{AB} = 2$.
	}
	\label{supp:fig:flux:W20}
	
	\subcaptionbox{%
		Min: \num{0.049}.
		Max: \num{3.732}.
	}{%
		\includegraphics{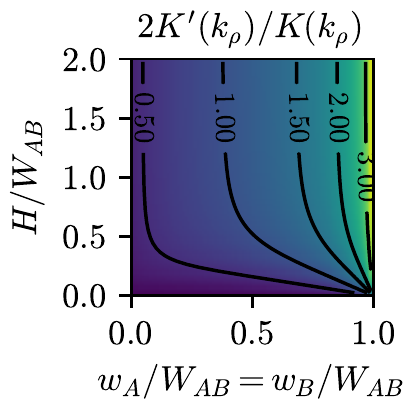}
	}%
	\subcaptionbox{%
		Min: \num{0.325}.
		Max: \num{3.200}.
	}{%
		\includegraphics{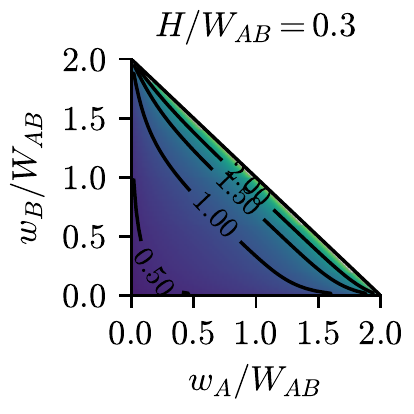}
	}%
	\subcaptionbox{%
		Min: \num{0.407}.
		Max: \num{3.744}.
	}{%
		\includegraphics{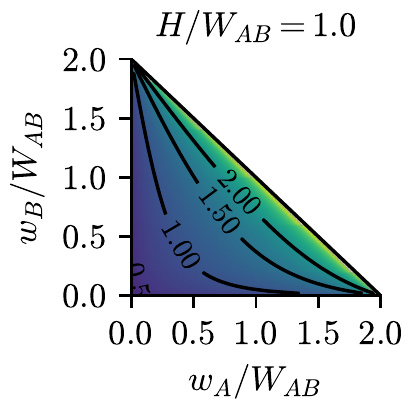}
	}%
	\caption{%
		Normalized flux per band $2K'(k_{\rho})/K(k_{\rho}) = (i_{A}/L)/(\gamma [u_{A} - u_{B}])$ as a function of the cell dimensions.
		Case of \emph{exterior cell} with $W/W_{AB} = 3$.
	}
	
	\subcaptionbox{%
		Min: \num{0.049}.
		Max: \num{3.732}.
	}{%
		\includegraphics{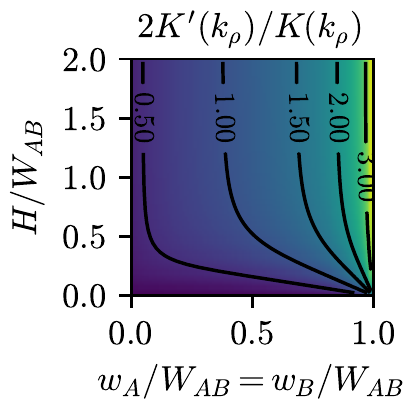}
	}%
	\subcaptionbox{%
		Min: \num{0.325}.
		Max: \num{3.200}.
	}{%
		\includegraphics{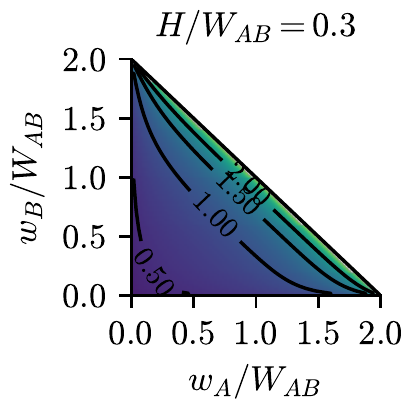}
	}%
	\subcaptionbox{%
		Min: \num{0.407}.
		Max: \num{3.744}.
	}{%
		\includegraphics{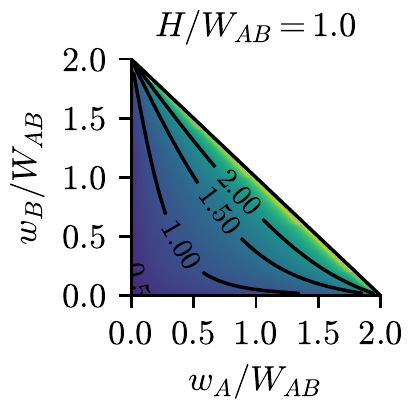}
	}%
	\caption{%
		Normalized flux per band $2K'(k_{\rho})/K(k_{\rho}) = (i_{A}/L)/(\gamma [u_{A} - u_{B}])$ as a function of the cell dimensions.
		Case of \emph{exterior cell} with $W/W_{AB} = 5$.
	}
\end{figure}

\begin{figure}
	\subcaptionbox{%
		Min: \num{0.000}.
		Max: \num{0.279}.
	}{%
		\includegraphics{fig-flux-error-W20-Hall}
	}%
	\subcaptionbox{%
		Min: \num{0.000}.
		Max: \num{0.522}.
	}{%
		\includegraphics{fig-flux-error-W20-H03}
	}%
	\subcaptionbox{%
		Min: \num{0.000}.
		Max: \num{0.619}.
	}{%
		\includegraphics{fig-flux-error-W20-H10}
	}%
	\caption{%
		Relative error of flux for an \emph{exterior cell} with respect to that of an \emph{interior cell} $i_{A}^\text{ext}/i_{A}^\text{int} - 1$ as a function of the cell dimensions.
		Case of \emph{exterior cell} with $W/W_{AB} = 2$.
	}
	
	\subcaptionbox{%
		Min: \num{0.000}.
		Max: \num{0.281}.
	}{%
		\includegraphics{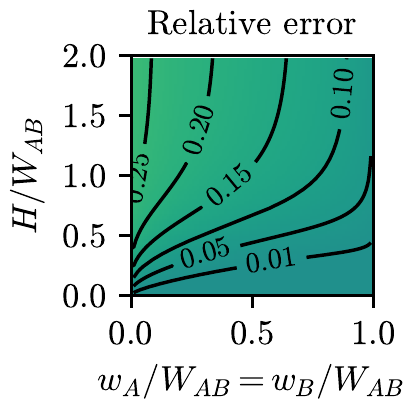}
	}%
	\subcaptionbox{%
		Min: \num{0.000}.
		Max: \num{0.522}.
	}{%
		\includegraphics{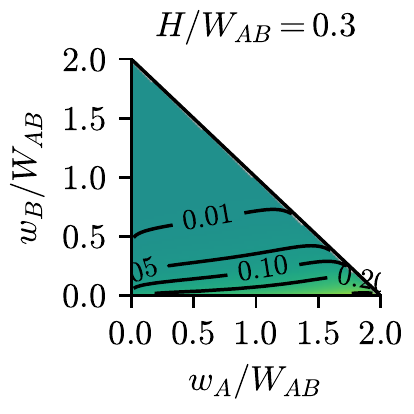}
	}%
	\subcaptionbox{%
		Min: \num{0.000}.
		Max: \num{0.619}.
	}{%
		\includegraphics{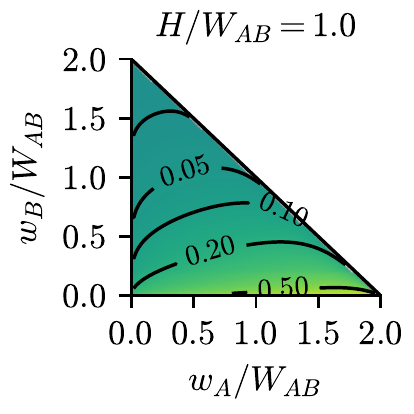}
	}%
	\caption{%
		Relative error of flux for an \emph{exterior cell} with respect to that of an \emph{interior cell} $i_{A}^\text{ext}/i_{A}^\text{int} - 1$ as a function of the cell dimensions.
		Case of \emph{exterior cell} with $W/W_{AB} = 3$.
	}
	
	\subcaptionbox{%
		Min: \num{0.000}.
		Max: \num{0.281}.
	}{%
		\includegraphics{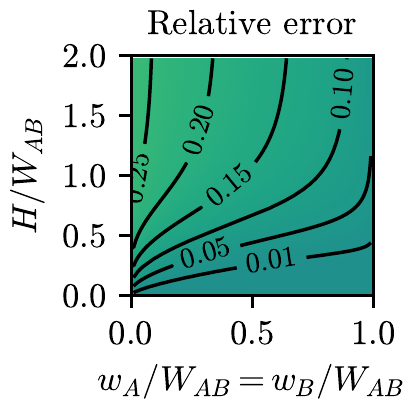}
	}%
	\subcaptionbox{%
		Min: \num{0.000}.
		Max: \num{0.522}.
	}{%
		\includegraphics{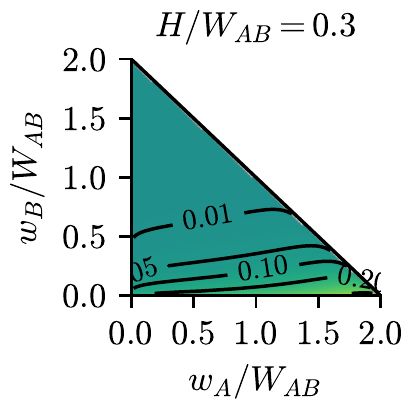}
	}%
	\subcaptionbox{%
		Min: \num{0.000}.
		Max: \num{0.619}.
	}{%
		\includegraphics{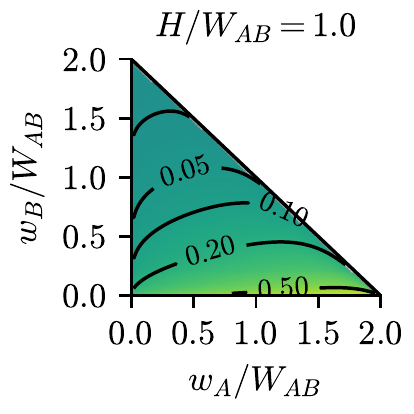}
	}%
	\caption{%
		Relative error of flux for an \emph{exterior cell} with respect to that of an \emph{interior cell} $i_{A}^\text{ext}/i_{A}^\text{int} - 1$ as a function of the cell dimensions.
		Case of \emph{exterior cell} with $W/W_{AB} = 5$.
	}
	\label{supp:fig:flux:error:W50}
\end{figure}

The normalized flux per band for \emph{interior} and \emph{exterior cells}
\begin{equation}
	\frac{i_{A}^\text{int}/L}{\gamma [u_{A} - u_{B}]} = 2\frac{K'(k_{\rho})}{K(k_{\rho})} \bigg|_\text{int},
	\quad
	\frac{i_{A}^\text{ext}/L}{\gamma [u_{A} - u_{B}]} = 2\frac{K'(k_{\rho})}{K(k_{\rho})} \bigg|_\text{ext}
\end{equation}
and the relative error of the flux for an \emph{exterior cell} with respect to that of an \emph{interior cell}
\begin{equation}
	\frac{i_{A}^\text{ext}}{i_{A}^\text{int}} - 1 =
	\frac{
		2 K'(k_{\rho})/K(k_{\rho})|_\text{ext}
	}{
		2 K'(k_{\rho})/K(k_{\rho})|_\text{int}
	} - 1
\end{equation}
as functions of their geometrical parameters ($w_{A}/W_{AB}$, $w_{B}/W_{AB}$, $H/W_{AB}$ and $W/W_{AB}$) are difficult to visualize graphically, since they corresponds to three- and four-dimensional scalar fields for the cases of \emph{interior} and \emph{exterior cells} respectively.

Therefore, \figs\ \ref{supp:fig:flux:W20} to \ref{supp:fig:flux:error:W50} (as well as \figs\ \ref{main:fig:flux} and \ref{main:fig:flux:cells_error} at the \emph{main text}) show two-dimensional slices and contour lines of $2K'(k_{\rho})/K(k_{\rho})$ and $i_{A}^\text{ext}/i_{A}^\text{int} - 1$, as functions of the relative dimensions of the cell.
These slices were chosen such that they provide enough information to understand how $2K'(k_{\rho})/K(k_{\rho})$ and $i_{A}^\text{ext}/i_{A}^\text{int} - 1$ behave for different domain dimensions.

\section{Approximations for the interior cell}
\label{supp:sec:approx}

Calculating the flux per band at an interior cell requires the evaluation of $K'(k)/K(k)$ and its modulus; \eqns\ \eqref{supp:eqn:qr-qr'}, \eqref{supp:eqn:v:AB} and \ref{supp:eqn:krho-krho'}; which can be done by using specialized software like
\href{https://www.python.org/}{\textsf{Python}},
\href{https://www.sagemath.org/}{\textsf{SageMath}}, \textsc{Matlab} and
\textsf{Mathematica} among others.
However, an expression for the flux involving simpler elementary functions is highly desirable, since this would enable the use of any general purpose software or calculator.

One convenient way to find such approximations is by using the \emph{nome function} $q = Q(k)$, shown in Eq. \eqref{supp:eqn:nomo}, as a way to compute the ratio%
\footnote{
	this ratio is closely related to the \emph{lattice parameter} $\tau = \bm{i} K'(k)/K(k)$ and the \emph{nome} $q = \exp(\bm{i} \pi \tau)$ 	\citex{\S\dlmf{20.}{1}, \S\dlmf{22.}{1}, and \eqns\ (\dlmf[E]{22.2.}{1}) and (\dlmf[E]{22.2.}{12})}{dlmf}.
} $K'(k)/K(k)$
\begin{equation*}
	\frac{K'(k)}{K(k)} = \frac{\ln Q(k)}{-\pi} = \frac{-\pi}{\ln Q(k')}
\end{equation*}
This is because the Taylor expansion of the nome function $Q(k)$ is known and converges relatively fast \cite[below \eqn\ (12)]{Kneser1927} \citex{\eqn\ (\dlmf[E]{19.5.}{5})}{dlmf}
\begin{equation}
	\label{supp:eqn:nomo:taylor}
	Q(k) =
	\frac{k^{2}}{16} + 8\left( \frac{k^{2}}{16} \right)^{2}
	+  84 \left( \frac{k^{2}}{16} \right)^{3}
	+ 992 \left( \frac{k^{2}}{16} \right)^{4} + O(k^{10})
\end{equation}
Therefore, for a sufficiently small modulus $k$ or $k'$, the ratio $K'(k)/K(k)$ can be approximated with enough accuracy by using only the first term of the previous series
\begin{subequations}
	\label{supp:eqn:K'K:approx}
	\begin{align}
		\frac{K'(k)}{K(k)} \bigg|_{k \approx 0}
		&=  -\frac{1}{\pi} \ln Q(k) \bigg|_{k \approx 0}
		\approx -\frac{1}{\pi} \ln\del{ \frac{k^{2}}{16} }
		\label{supp:eqn:K'K:k0}
		\\
		\frac{K'(k)}{K(k)} \bigg|_{k' \approx 0}
		&=  -\pi [\ln Q(k')]^{-1} \bigg|_{k' \approx 0}
		\approx -\pi \sbr{ \ln\del{ \frac{{k'}^{2}}{16} } }^{-1}
		\label{supp:eqn:K'K:k'0}
	\end{align}%
\end{subequations}%
as shown before in the literature \cite[above \eqn\ (32)]{Aoki1988dec}
\cite[\eqn\ (5)]{Morf2006may}.

To aid in the process of approximation, it is convenient to rewrite the modulus $k_{\rho}$ and its complement $k_{\rho}'$.
Since for an \emph{interior cell} we have
\begin{equation*}
	2w_{a} = w_{A},\quad 2w_{b} = w_{B}, \quad \text{and} \quad \bm{r}_{b} = W_{AB} = \bm{r}_{l} = W
\end{equation*}%
the points of interest of domain $\bm{v}$ in \eqns\ \eqref{supp:eqn:v:AB} reduce to
\begin{subequations}
	\begin{align}
		\bm{v}_{a} = -1, & \quad
		\bm{v}_{\alpha} = -\cd\del{ K(k_{r}) \frac{w_{A}}{W_{AB}}, k_{r} }
		\\
		\bm{v}_{b} = 1, & \quad
		\bm{v}_{\beta}
		= -\cd\del{ 2K(k_{r}) - \frac{w_{B}}{W_{AB}}, k_{r} }
		= \cd\del{ K(k_{r}) \frac{w_{B}}{W_{AB}}, k_{r} }
	\end{align}%
\end{subequations}
due to half period translation $\cd(\bm{u} + 2K(k), k) = -\cd(\bm{u}, k)$ \citex{Table \dlmf[T]{22.4.}{3}}{dlmf} and
due to double rotation of the argument $\cd(-\bm{u}, k) = \cd(\bm{i}^{2} \bm{u}, k) = \cd(\bm{u}, k)$ \citex{Table \dlmf[T]{22.6.}{1}}{dlmf}.
This also reduces $q_{r}$ and $q_{r}'$ in \eqns\ \eqref{supp:eqn:qr-qr'} to
\begin{equation}
	\label{supp:eqn:qr-qr':int}
	q_{r} = Q(k_{r}) = \exp\del{ -\pi \frac{2H}{W_{AB}} }, \quad
	q'_{r} = Q(k'_{r}) = \exp\del{ -\pi \frac{W_{AB}}{2H} }
\end{equation}
and the moduli $k_{\rho}$ and $k_{\rho}'$ in \eqns\ \eqref{supp:eqn:krho-krho'} to
\begin{subequations}
	\begin{equation}
		k_{\rho}^{2} = \frac{
			\displaystyle
			2 \sbr{
				\cd\del{ K(k_{r}) \frac{w_{A}}{W_{AB}},\, k_{r} } +
				\cd\del{ K(k_{r}) \frac{w_{B}}{W_{AB}},\, k_{r} }
			}
		}{
			\displaystyle
			\sbr{ 1 + \cd\del{ K(k_{r}) \frac{w_{A}}{W_{AB}},\, k_{r} } }
			\sbr{ 1 + \cd\del{ K(k_{r}) \frac{w_{B}}{W_{AB}},\, k_{r} } }
		}
	\end{equation}
	\begin{equation}
		{k_{\rho}'}^{2} = \frac{
			\displaystyle
			\sbr{ 1 - \cd\del{ K(k_{r}) \frac{w_{A}}{W_{AB}},\, k_{r} } }
		}{
			\displaystyle
			\sbr{ 1 + \cd\del{ K(k_{r}) \frac{w_{A}}{W_{AB}},\, k_{r} } }
		}
		\frac{
			\displaystyle
			\sbr{ 1 - \cd\del{ K(k_{r}) \frac{w_{B}}{W_{AB}},\, k_{r} } }
		}{
			\displaystyle
			\sbr{ 1 + \cd\del{ K(k_{r}) \frac{w_{B}}{W_{AB}},\, k_{r} } }
		}
	\end{equation}%
\end{subequations}%
This further leads to an alternative representation for $k_{\rho}$, which depends on the gap $g = W_{AB} - w_{A} = W_{AB} - w_{B}$ between bands of equal width $w_{A} = w_{B}$
\begin{subequations}
	\begin{equation}
		k_{\rho}^{2} = \frac{
			\displaystyle
			4\, \sn\del{ K(k_{r}) \frac{g}{W_{AB}},\, k_{r} }
		}{
			\displaystyle
			\sbr{ 1 + \sn\del{ K(k_{r}) \frac{g}{W_{AB}},\, k_{r} } }^{2}
		}
	\end{equation}%
	due to quarter period translation $\cd(\bm{u} + K(k),k) = -\sn(\bm{u}, k)$ \citex{\tab\ \dlmf[T]{22.4.}{3}}{dlmf} and doble rotation of the argument $\sn(-\bm{u}, k) = \sn(\bm{i}^{2}\bm{u}, k) = -\sn(\bm{u}, k)$ \citex{\tab\ \dlmf[T]{22.6.}{1}}{dlmf}.
	Also $k_{\rho}'$ has an alternative representation, which can be expressed using different band widths
	\begin{equation}
		{k_{\rho}'}^{2} = {k_{r}'}^{4} \frac{
			\displaystyle
			\sd\del{ \frac{K(k_{r})}{2} \frac{w_{A}}{W_{AB}},\, k_{r} }^{2}
		}{
			\displaystyle
			\cn\del{ \frac{K(k_{r})}{2} \frac{w_{A}}{W_{AB}},\, k_{r} }^{2}
		}
		\frac{
			\displaystyle
			\sd\del{ \frac{K(k_{r})}{2} \frac{w_{B}}{W_{AB}},\, k_{r} }^{2}
		}{
			\displaystyle
			\cn\del{ \frac{K(k_{r})}{2} \frac{w_{B}}{W_{AB}},\, k_{r} }^{2}
		}
	\end{equation}%
\end{subequations}%
due to $[1 - \cd(2\bm{u},k)]/[1 + \cd(2\bm{u},k)] = {k'}^{2} \sd(\bm{u},k)^{2}/\cn(\bm{u},k)^{2}$ \cite[\eqns\ (1.10) and (4.1)]{Carlson2004nov}.
Here the special functions $\sn$, $\sd$, $\cn$ and $\cd$ correspond to \emph{Jacobian elliptic functions} analogous to their circular counterparts $\sin$ and $\cos$ \citex{\S\dlmf{22.}{2}}{dlmf}.

Finally, with this alternative representations for $k_{\rho}$ and $k_{\rho}'$, we can obtain approximations based on elementary functions for the cases of tall and shallow domains.

\subsection{Approximations for tall domains}

Consider the following approximations for sufficiently small nome $q = Q(k)$
\begin{subequations}
	\label{supp:eqn:jacobi:q0}
	\begin{align}
		\lim_{q \to 0^{+}} \sn(\bm{u}, k)
		= \lim_{q \to 0^{+}} \sd(\bm{u}, k)
		&= \sin\del{ \frac{\pi}{2} \frac{\bm{u}}{K(k)} }
		\\
		\lim_{q \to 0^{+}} \cn(\bm{u}, k)
		= \lim_{q \to 0^{+}} \cd(\bm{u}, k)
		&= \cos\del{ \frac{\pi}{2} \frac{\bm{u}}{K(k)} }
		\\
		\lim_{q \to 0^{+}} \dn(\bm{u}, k)
		= \lim_{q \to 0^{+}} \nd(\bm{u}, k)
		&= 1
	\end{align}%
\end{subequations}%
\citex{\eqn\ (\dlmf[E]{19.6.}{1}) and \tab\ \dlmf[T]{22.5.}{3}}{dlmf} or \cite[\eqns\ (10)]{Fenton1982jul}.

In case of tall domains ($H$ is large) $q_{r} \to 0^{+}$ and $k_{r} \to 0^{+}$, therefore we obtain
\begin{subequations}
	\label{supp:eqn:moduli:Hinf}
	\begin{align}
	\label{supp:eqn:krho:Hinf}
	\bigg. k_{\rho}^{2} \bigg|_{
		\scriptsize \shortstack{
			$H \to +\infty$ \\
			$g \approx 0$
		}
	}
	&= \frac{
		\displaystyle
		4 \sin\del{ \frac{\pi}{2} \frac{g}{W_{AB}} }
	}{
		\displaystyle
		\sbr{ 1 + \sin\del{ \frac{\pi}{2} \frac{g}{W_{AB}} } }^{2}
	}
	\approx 4 \del{ \frac{\pi}{2}\frac{g}{W_{AB}} }
	\\
	\label{supp:eqn:krho':Hinf}
	\bigg. {k_{\rho}'}^{2} \bigg|_{
		\scriptsize \shortstack{
			$H \to +\infty$ \\
			$w_{A} \approx 0$ \\
			$w_{B} \approx 0$
		}
	}
	&=
	\tan\del{ \frac{\pi}{4} \frac{w_{A}}{W_{AB}} }^{2}
	\tan\del{ \frac{\pi}{4} \frac{w_{B}}{W_{AB}} }^{2}
	\approx
	\del{ \frac{\pi}{4} \frac{w_{A}}{W_{AB}} }^{2}
	\del{ \frac{\pi}{4} \frac{w_{B}}{W_{AB}} }^{2}
	\end{align}%
\end{subequations}%

\subsection{Approximations for shallow domains}

Consider the following approximations when the associated nome $q' = Q(k')$ is sufficiently small \cite[\eqns\ (11)]{Fenton1982jul}
\begin{subequations}
	\label{supp:eqn:jacobi:q'0}
	\begin{align}
		\sn(\bm{u}, k) \bigg|_{q' \approx 0} &\approx
		(k^{2})^{-1/4} \tanh\del{ \frac{\pi}{2} \frac{\bm{u}}{K'(k)} }
		\\
		\cn(\bm{u}, k) \bigg|_{q' \approx 0} &\approx
		\frac{1}{2} \del{ \frac{{k'}^{2}}{k^{2} q'} }^{1/4}
		\sech\del{ \frac{\pi}{2} \frac{\bm{u}}{K'(k)} }
		\\
		\dn(\bm{u}, k) \bigg|_{q' \approx 0} &\approx
		\frac{1}{2} \del{ \frac{{k'}^{2}}{q'} }^{1/4}
		\sech\del{ \frac{\pi}{2} \frac{\bm{u}}{K'(k)} }
	\end{align}%
	also of the subsidiary function $\sd = \sn/\dn$
	\begin{align}
		\sd(\bm{u}, k) \bigg|_{q' \approx 0} &\approx
		2 \del{ \frac{q'}{k^{2} {k'}^{2}} }^{1/4}
		\sinh\del{ \frac{\pi}{2} \frac{\bm{u}}{K'(k)} }
	\end{align}%
	and of the modulus $(k^{2})^{1/4}$ \cite[begining of \S7]{Fenton1982jul}
	\begin{align}
		(k^{2})^{1/4} \bigg|_{q' \approx 0} &\approx
		\frac{1 - 2q'}{1 + 2q'} =
		\tanh\del{ \frac{\pi}{2} \frac{K(k)}{K'(k)} - \frac{1}{2} \ln 2 }
	\end{align}%
\end{subequations}%
where $q' = Q(k') = \exp(-\pi K(k)/K(k'))$.

In case of shallow domains ($H$ is small) $q_{r}' \to 0^{+}$ and $k_{r}' \to 0^{+}$, which leads to
\begin{subequations}
	\label{supp:eqn:moduli:H0}
	\begin{align}
		\label{supp:eqn:krho:H0}
		k_{\rho}^{2} \bigg|_{\scriptsize \shortstack{$q_{r}' \approx 0$ \\ $g \approx 0$}}
		&\approx
		4\sn\del{ K(k_{r}) \frac{g}{W_{AB}}, k_{r} } \bigg|_{q_{r}' \approx 0}
		\approx
		\frac{
			\displaystyle
			4\tanh\del{
				\frac{\pi}{2} \frac{K(k_{r})}{K'(k_{r})} \frac{g}{W_{AB}}
			}
		}{
			\displaystyle
			\tanh\del{
				\frac{\pi}{2} \frac{K(k_{r})}{K'(k_{r})} - \ln\sqrt{2}
			}
		}
		\\
		\label{supp:eqn:krho':H0}
		{k_{\rho}'}^{2} \bigg|_{q_{r}' \approx 0}
		&\approx
		16 {q_{r}'}^{2}\,
		\sinh\del{ \frac{\pi}{2} \frac{K(k_{r})}{K'(k_{r})} \frac{w_{A}}{W_{AB}} }^{2}
		\sinh\del{ \frac{\pi}{2} \frac{K(k_{r})}{K'(k_{r})} \frac{w_{B}}{W_{AB}} }^{2}
	\end{align}%
\end{subequations}%
where the last equality was obtained through the identity $2 \sinh(\bm{u}) \cosh(\bm{u}) = \sinh(2\bm{u})$, as shown below for $E \in \{A, B\}$
\begin{equation}
	{k_{r}'}^{2}\,
	\frac{
		\displaystyle
		\sd\del{ \frac{K(k_{r})}{2} \frac{w_{E}}{W_{AB}}, k_{r} }^{2}
	}{
		\displaystyle
		\cn\del{ \frac{K(k_{r})}{2} \frac{w_{E}}{W_{AB}}, k_{r} }^{2}
	}
	\approx
	16 {q_{r}'}\,
	\frac{
		\displaystyle
		\sinh\del{
			\frac{\pi}{4} \frac{K(k_{r})}{K'(k_{r})} \frac{w_{E}}{W_{AB}}
		}^{2}
	}{
		\displaystyle
		\sech\del{
			\frac{\pi}{4}\frac{K(k_{r})}{K'(k_{r})} \frac{w_{E}}{W_{AB}}
		}^{2}
	}
	=
	4 {q_{r}'}\,
	\sinh\del{ \frac{\pi}{2} \frac{K(k_{r})}{K'(k_{r})} \frac{w_{E}}{W_{AB}} }^{2}
\end{equation}
Here $K'(k_{r})/K(k_{r}) = 2H/W_{AB}$ was obtained from $q_{r}' = Q(k_{r}') = \exp(-\pi K(k_{r})/K(k_{r}')) = \exp(-\pi W_{AB}/2H)$, due to \eqns\ \eqref{supp:eqn:nomo} and \eqref{supp:eqn:qr-qr':int}.

\section{Comparison with previous results}
\label{supp:sec:comparison}

In this section, simulations and theoretical expressions from the literature are written using the notation of this work, in order to compare against the normalized flux per band in \eqns\ \eqref{main:eqn:flux:band} of the \emph{main text}
\begin{equation}
	\frac{i_{A}/L}{\gamma [u_{A} - u_{B}]} = 2\frac{K'(k_{\rho})}{K(k_{\rho})}
\end{equation}
Note that all these results from the literature consider arrays with bands of equal width $w_{E} := w_{A} = w_{B}$ and an \emph{interior cell}.

\begin{table}
	\centering
	\begin{tabular}{lrrrr}
\toprule
{} & \multicolumn{2}{c}{$w_{E}/g = 0.5$} & \multicolumn{2}{c}{$w_{E}/g = 1$} \\
\cmidrule(r){2-3} \cmidrule(r){4-5}
{} & $w_{E}/H$ & $|G_{ss}|$ & $w_{E}/H$ & $|G_{ss}|$ \\ 
\midrule
0  &  0.259325 &  0.779973 &  0.259325 &  0.995728 \\
1  &  0.490231 &  0.766088 &  0.490231 &  0.989319 \\
2  &  0.724689 &  0.712684 &  0.724689 &  0.972230 \\
3  &  0.959147 &  0.642190 &  0.959147 &  0.931642 \\
4  &  1.190053 &  0.572764 &  1.193606 &  0.875033 \\
5  &  1.420959 &  0.514019 &  1.424512 &  0.812016 \\
6  &  1.655417 &  0.462750 &  1.658970 &  0.751135 \\
7  &  1.889876 &  0.418959 &  1.893428 &  0.696662 \\
8  &  2.124334 &  0.381575 &  2.127886 &  0.649666 \\
9  &  2.355240 &  0.351669 &  2.358792 &  0.606943 \\
10 &  2.589698 &  0.326035 &  2.593250 &  0.568491 \\
11 &  2.824156 &  0.302537 &  2.827709 &  0.534312 \\
12 &  3.058615 &  0.283311 &  3.062167 &  0.502270 \\
13 &  3.289520 &  0.265154 &  3.293073 &  0.474499 \\
14 &  3.523979 &  0.250200 &  3.527531 &  0.447797 \\
15 &  3.758437 &  0.236315 &  3.761989 &  0.426435 \\
\bottomrule
\end{tabular}

	\caption{
		Normalized faradaic currents $|G_{ss}| = 2K'(k_{\rho})/K(k_{\rho})$ for an \emph{interior cell} with bands of equal width $w_{E} = w_{A} = w_{B}$ \cite[\fig\ 7a]{Strutwolf2005feb}.
		Data obtained by simulating for different values of $w_{E}/H$ and $w_{E}/g \in \cbr{\num{0,5}, 1}$.
	}
	\label{supp:tab:itau:Strutwolf2005}
\end{table}

\tab\ \ref{supp:tab:itau:Strutwolf2005} shows simulations obtained in \cite[Fig. 7a]{Strutwolf2005feb} for the normalized faradaic current in an \emph{interior cell}.
The expression \cite[$|G_{ss}|$ corresponds to $|G_{g}| = |G_{c}|$ in \S2.2 and \eqn\ (3) at steady state]{Strutwolf2005feb}, which was rewritten using the notation of this text
\begin{equation}
	|G_{ss}| = \left| \frac{
		N_{A} i_{A}
	}{
		L N_{A} \underbrace{n_{e} D F}_{\gamma}
		[\underbrace{c_{b}}_{u_{A}} - \underbrace{0}_{u_{B}}]
	} \right|
	= 2\frac{K'(k_{\rho})}{K(k_{\rho})}
\end{equation}
where $n_{e}$ is the number of electrons exchanged in the Faradaic process, $D$ is the diffusion coefficient of the redox species, $F$ is the Faraday's constant and $c_{b}$ corresponds to the bulk concentration of species in the electrochemical cell.
These simulated results correspond to the case of a process of limiting current (with counter electrode external to the IDA), where the concentration at the bands of one array equals $c_{b}$, while the concentration at the bands of the other array equals zero.
For plotting this data in \fig\ \ref{main:fig:flux:comparison} at the \emph{main text}, the horizontal variable \cite[$w_{e}/h_{c}$ in \fig\ 7a]{Strutwolf2005feb} was also rewritten according to the notation of this text, that is $w_{E}/H$ in \tab\ \ref{supp:tab:itau:Strutwolf2005}, and later related to $H/W$ by
\begin{equation}
	\frac{H}{W} = \left[
	\frac{w_{E}}{H} \cdot
	\left( \frac{g}{w_{E}} + \frac{w_{E}}{w_{E}} \right)
	\right]^{-1}
\end{equation}
where $g$ is the gap between electrodes such that $W_{AB} = w_{E} + g$.
Only the data for the case $w_{E} = g$ ($w_{E}/W_{AB} =\num{0,5}$) was plotted in \fig\ \ref{main:fig:flux:comparison} at the \emph{main text}.

\begin{table}
	\centering
	\begin{tabular}{rrrrr}
\toprule
{} & \multicolumn{4}{c}{$w_{E}/W_{AB}$} \\
\cmidrule(lr){2-5}
$H/W_{AB}$ &      0.2 &      0.4 &      0.6 &      0.8 \\
\midrule
$0.2/\pi$ &  0.01414 &  0.01845 &  0.02650 &  0.04729 \\
$0.4/\pi$ &  0.02652 &  0.03458 &  0.04817 &  0.07909 \\
$0.6/\pi$ &  0.03628 &  0.04792 &  0.06546 &  0.10152 \\
$0.8/\pi$ &  0.04353 &  0.05844 &  0.07902 &  0.11807 \\
$1/\pi$   &  0.04873 &  0.06636 &  0.08932 &  0.13031 \\
$2/\pi$   &  0.05864 &  0.08236 &  0.11050 &  0.15522 \\
$3/\pi$   &  0.06007 &  0.08475 &  0.11373 &  0.15926 \\
$4/\pi$   &  0.06026 &  0.08508 &  0.11417 &  0.15954 \\
$5/\pi$   &  0.06029 &  0.08513 &  0.11424 &  0.15962 \\
\bottomrule
\end{tabular}

	\caption{
		Normalized faradaic currents $(2/\pi^{2})\, K'(k_{\rho})/K(k_{\rho})$ for bands of equal width $w_{E} = w_{A} = w_{B}$ \cite[\fig\ 7a]{GuajardoYevenes2013sep}.
		Data obtained by simulating exhaustively for different values of $H/W_{AB}$ and $w_{E}/W_{AB}$.
	}
	\label{supp:tab:itau:GuajardoYevenes2013}
\end{table}

\tab\ \ref{supp:tab:itau:GuajardoYevenes2013} shows exhaustive simulations obtained in  \cite[Fig. 7a]{GuajardoYevenes2013sep} for the normalized faradaic current in an \emph{interior cell}.
The expression for the average flux density \cite[$\bar{\phi}_{\lambda}^{\lim}$]{GuajardoYevenes2013sep} was related to the current per array $N_{A} i_{A} = N_{A} L w_{E}\, F n_{e} \bar{\phi}_{\lambda}^{\lim}$ \cite[\eqn\ (12)]{GuajardoYevenes2013sep}, and then replaced in the expression of \cite[Fig. 7a]{GuajardoYevenes2013sep}
\begin{equation}
	\left| \frac{
		\bar{\phi}_{\lambda}^{\lim} w_{E}/2
	}{
		\pi^{2} D \bar{c}_{\ell,0}
	} \right|
	= \frac{1}{\pi^{2}}
	\left| \frac{
		i_{A}/L
	}{
		\underbrace{F n_{e} D}_{\gamma}
		[\underbrace{2 \bar{c}_{\ell,0}}_{u_{A}} - \underbrace{0}_{u_{B}}]
	} \right|
	= \frac{1}{\pi^{2}} \cdot 2\frac{K'(k_{\rho})}{K(k_{\rho})}
\end{equation}
showing that the normalized currents in \cite{GuajardoYevenes2013sep} and in this work are related by a factor of $\pi^{2} \approx 10$.
Here $\bar{c}_{\ell,0}$ is the smallest initial concentration between both redox species in the electrochemical cell.
These simulated results correspond to the case of a process of limiting current (with one of the arrays acting as counter electrode), where the concentration at the bands of one array equals $2 \bar{c}_{\ell,0}$, while the concentration at the bands of the other array equals zero \cite[\fig\ 2]{GuajardoYevenes2013sep}.
See \fig\ \ref{main:fig:flux:comparison} at the \emph{main text} for a graphical representation of \tab\ \ref{supp:tab:itau:GuajardoYevenes2013}.

The expression \cite[$C_{I}$ in \eqn\ (13)]{Igreja2004may} shows the capacitance of what corresponds to half \emph{interior cell} in this work.
In \cite{Igreja2004may} it is assumed that the potentials applied at bands of each array are symmetric ($u_{A} = V$ and $u_{B} = -V$).
Despite this, the difference of potentials inside half \emph{interior cell} is only $V$, because this contains only half band of one array.
Also, since the charge (flux of electric displacement) at one band corresponds to $i_{A}$, then in half \emph{interior cell} we have a charge of $i_{A}/2$.
Therefore, the normalized capacitance in half \emph{interior cell} can be rewritten using the notation of this text as
\begin{equation}
	\frac{C_{I}}{\epsilon L}
	= \frac{1}{\epsilon L} \frac{i_{A}/2}{V}
	= \frac{i_{A}/L}{
		\underbrace{\epsilon}_{\gamma}
		[\underbrace{V}_{u_{A}} - \underbrace{-V}_{u_{B}}]
	}
	= 2\frac{K'(k_{\rho})}{K(k_{\rho})}
\end{equation}
where $\epsilon = \epsilon_{0} \epsilon_{r}$ is the absolute permittivity of the dielectric.
See \fig\ \ref{main:fig:flux:comparison} at the \emph{main text} for a comparison between \cite[$C_{I}/(\epsilon L)$ in \eqn\ (13)]{Igreja2004may} with $2K'(k_{\rho})/K(k_{\rho})$. 

\section{Regions of validity for approximated flux}
\label{supp:sec:regions}

The relative error between the approximated and exact fluxes ($N_{A} \tilde{i}_{A}$ and $N_{A} i_{A}$) for an \emph{ideal domain} corresponds to that between their normalized counterparts
\begin{equation}
	\text{relative error} =
	\frac{N_{A} \tilde{i}_{A}}{N_{A} i_{A}} - 1 =
	\left[2 \frac{K'(k_{\rho})}{K(k_{\rho})} \right]_{\text{app.}}
	\left[2 \frac{K'(k_{\rho})}{K(k_{\rho})} \right]^{-1} - 1
\end{equation}
This was used to find the combinations of band widths and domain height that produce a relative error less than $\num{+-0,05} = \SI{+-5}{\percent}$, thus defining regions of parameters where the approximations are valid.
The search for these regions of validity was done graphically, and therefore, it was more convenient to consider bands of equal width $w_{E} := w_{A} = w_{B}$, such that the resulting plots were two-dimensional.

When expressing the ratio $K'(k)/K(k)$ in terms of the nome function $Q(k)$, or equivalently $Q(k')$ due to \eqn\ \eqref{supp:eqn:nomo}, one can realize that its approximations are due to a two-step approximation process
\begin{equation}
	\left[\frac{K'(k_{\rho})}{K(k_{\rho})} \right]_{\text{app.}} :=
	\left\{\begin{array}{ll}
	\frac{
		\displaystyle \ln\tilde{Q}(\tilde{k}_{\rho})
	}{\displaystyle -\pi}, & \text{for } k_{\rho} \approx 0
	\\[1em]
	\frac{\displaystyle -\pi}{
		\displaystyle \ln\tilde{Q}(\tilde{k}_{\rho}')
	}, & \text{for } k_{\rho}' \approx 0
	\end{array}\right.
\end{equation}
First, approximation of the nome function due to \eqn\ \eqref{supp:eqn:nomo:taylor} by $\tilde{Q}(k) := k^{2}/16$, and later the approximations $\tilde{k}_{\rho}$ and $\tilde{k}_{\rho}'$ of the moduli, which are given in \eqns\ \eqref{supp:eqn:moduli:Hinf} and \eqref{supp:eqn:moduli:H0}.

\begin{figure}[t]
	\centering
	\includegraphics{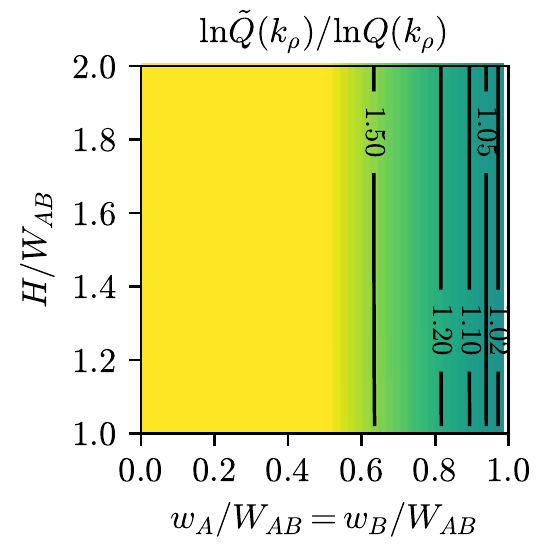} 
	\includegraphics{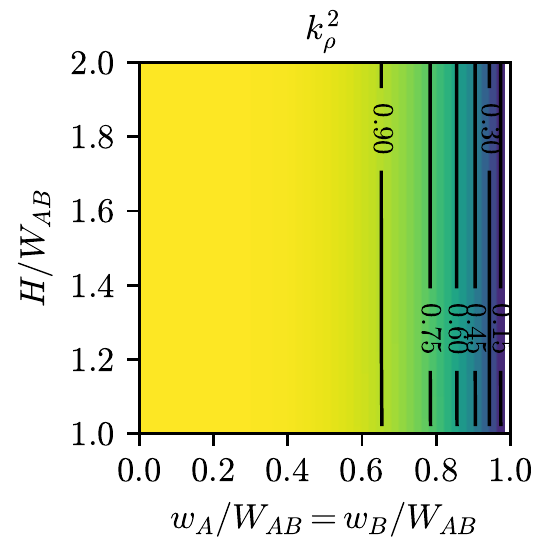}
	\\
	\includegraphics{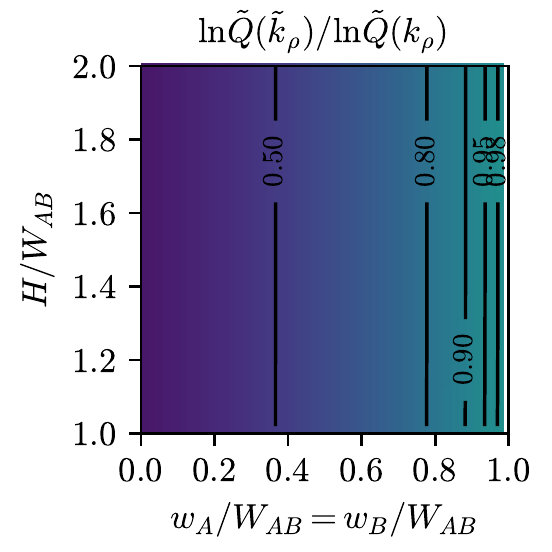} 
	\includegraphics{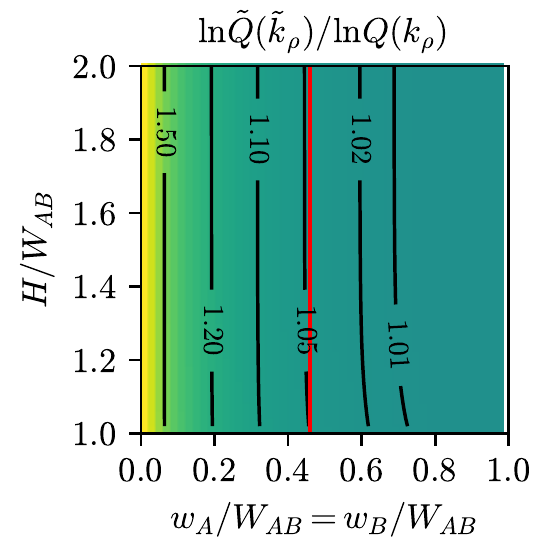}
	\caption{
		Propagation of the ratio of approximation $\ln\tilde{Q}(\tilde{k}_{\rho})/\ln Q(k_{\rho})$ for the case of tall domains and wide bands.
		Contour lines of selected values are shown in black.
		Line $w_{E}/W_{AB} = \num{0,46}$ is shown in red.
	}
	\label{supp:fig:ratios:tall:wide}
\end{figure}

Therefore, it is possible to analyze the propagation of the relative error, for $k_{\rho} \approx 0$, by looking at the following total ratio
\begin{equation}
	\label{supp:eqn:ratio:k_rho}
	\frac{N_{A} \tilde{i}_{A}}{N_{A} i_{A}} =
	\left[\frac{K'(k_{\rho})}{K(k_{\rho})} \right]_{\text{app.}}
	\left[\frac{K'(k_{\rho})}{K(k_{\rho})} \right]^{-1} =
	\frac{\ln\tilde{Q}(\tilde{k}_{\rho})}{\ln Q(k_{\rho})} =
	\frac{\ln\tilde{Q}(\tilde{k}_{\rho})}{\ln\tilde{Q}(k_{\rho})}
	\frac{\ln\tilde{Q}(k_{\rho})}{\ln Q(k_{\rho})}
\end{equation}
where a relative error of $\num{+-0,05} = \SI{+-5}{\percent}$ corresponds to the total ratios \num{0,95} and \num{1,05}.
The ratios $\ln\tilde{Q}(k_{\rho})/\ln Q(k_{\rho})$ and $\ln\tilde{Q}(\tilde{k}_{\rho})/\ln\tilde{Q}(k_{\rho})$ show the individual contributions of the approximated nome $\tilde{Q}$ and the approximated modulus $\tilde{k}_{\rho}$ to the total ratio $\ln\tilde{Q}(\tilde{k}_{\rho})/\ln Q(k_{\rho})$, and therefore, to the relative error.

This is shown in \fig\ \ref{supp:fig:ratios:tall:wide} for the case of tall domains and wide bands.
The first ratio $\ln\tilde{Q}(k_{\rho})/\ln Q(k_{\rho}) < \num{1,05}$ (relative error of \SI{5}{\percent}) for band widths $w_{E}/W_{AB} > \num{0,94}$ ($k_{\rho}^{2} \lesssim \num{0,316}$), which corresponds to very extreme cases of wide bands.
This agrees with the relative error of $\approx \SI{4}{\percent}$ in \cite[before \eqn\ (32)]{Aoki1988dec} when approximating only the nome function $\tilde{Q}(k_{\rho})$ for gap widths $g/W_{AB} < \num{0,059} \Leftrightarrow w_{E}/W_{AB} > \num{0,941}$ ($k_{\rho}^{2} < \num{0,310}$).
Fortunately, the contribution due to the approximation of the modulus $\tilde{k}_{\rho}$ in \eqn\ \eqref{supp:eqn:krho:Hinf} makes the second ratio $\ln\tilde{Q}(\tilde{k}_{\rho})/\ln\tilde{Q}(k_{\rho}) < 1$ in almost the whole domain $(w_{E}/W, H/W_{AB})$.
This helps to extend the region where the total ratio $\ln\tilde{Q}(\tilde{k}_{\rho})/\ln Q(k_{\rho}) < \num{1,05}$ (relative error less than \SI{5}{\percent}), which corresponds to more reasonable band widths $w_{E}/W_{AB} > \num{0,46}$.

\begin{figure}[t]
	\centering
	\includegraphics{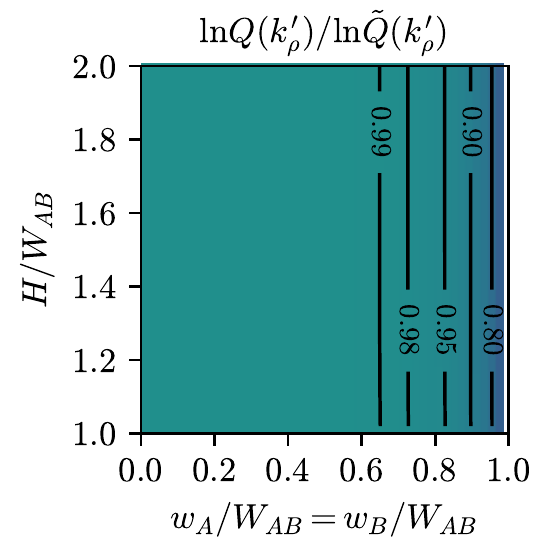} 
	\includegraphics{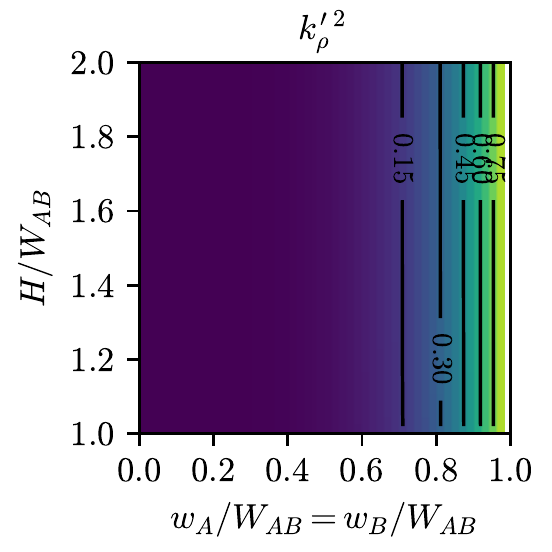}
	\\
	\includegraphics{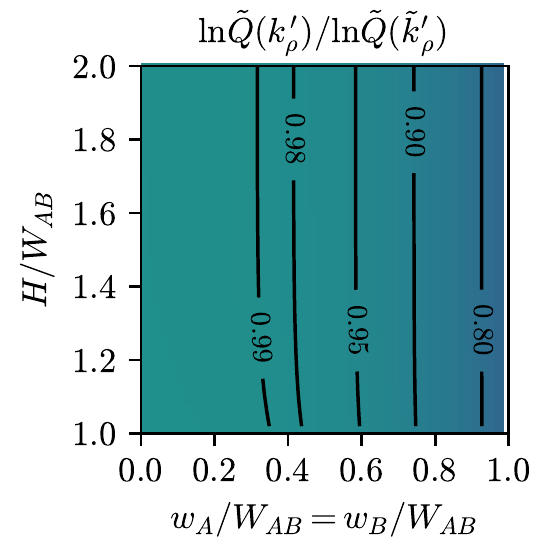} 
	\includegraphics{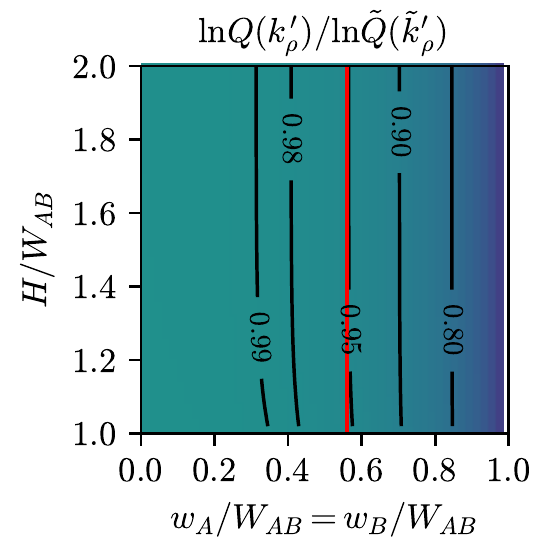}
	\caption{
		Propapation of the ratio of approximation $\ln\tilde{Q}(\tilde{k}_{\rho})/\ln Q(k_{\rho})$ for the case of tall domains and narrow electrodes.
		Contour lines of selected values are shown in black.
		Line $w_{E}/W_{AB} = \num{0,56}$ is shown in red.
	}
	\label{supp:fig:ratios:tall:narrow}
\end{figure}

A similar analysis of the propagation of the relative error can be done for $k_{\rho}' \approx 0$ when considering the following total ratio
\begin{equation}
	\label{supp:eqn:ratio:k'_rho}
	\frac{N_{A} \tilde{i}_{A}}{N_{A} i_{A}} =
	\left[\frac{K'(k_{\rho})}{K(k_{\rho})} \right]_{\text{app.}}
	\left[\frac{K'(k_{\rho})}{K(k_{\rho})} \right]^{-1} =
	\frac{\ln Q(k_{\rho}')}{\ln\tilde{Q}(\tilde{k}_{\rho}')} =
	\frac{\ln\tilde{Q}(k_{\rho}')}{\ln\tilde{Q}(\tilde{k}_{\rho}')}
	\frac{\ln Q(k_{\rho}')}{\ln\tilde{Q}(k_{\rho}')}
\end{equation}
where the ratios $\ln Q(k_{\rho}')/\ln\tilde{Q}(k_{\rho}')$ and $\ln\tilde{Q}(k_{\rho}')/\ln\tilde{Q}(\tilde{k}_{\rho}')$ show the individual contributions of the approximated nome $\tilde{Q}$ and the approximated complementary modulus $\tilde{k}_{\rho}'$ to the total ratio $\ln Q(k_{\rho}')/\ln\tilde{Q}(\tilde{k}_{\rho}')$, and therefore, to the relative error.

\begin{figure}[t]
	\centering
	\includegraphics{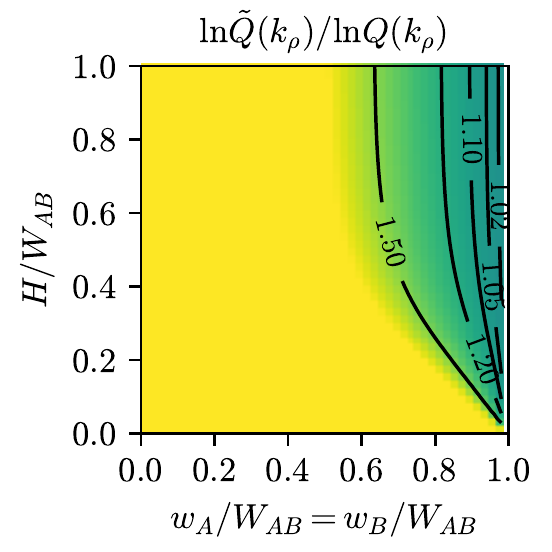} 
	\includegraphics{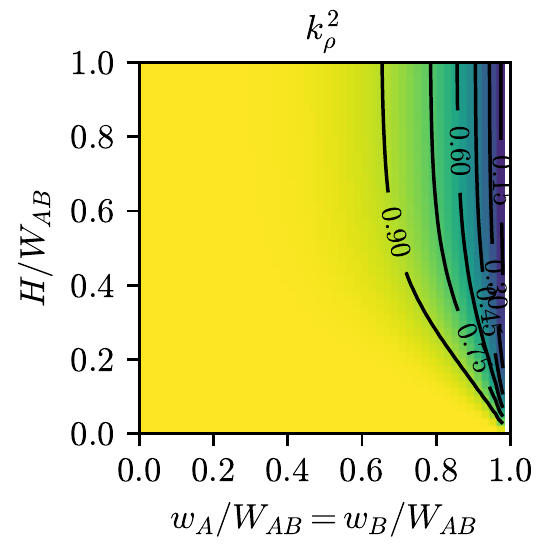}
	\\
	\includegraphics{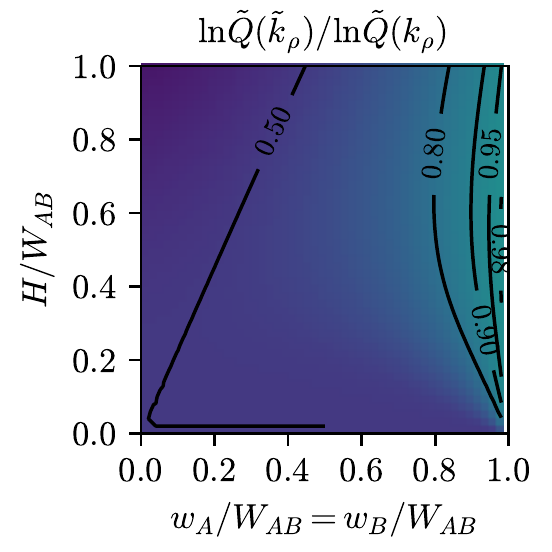} 
	\includegraphics{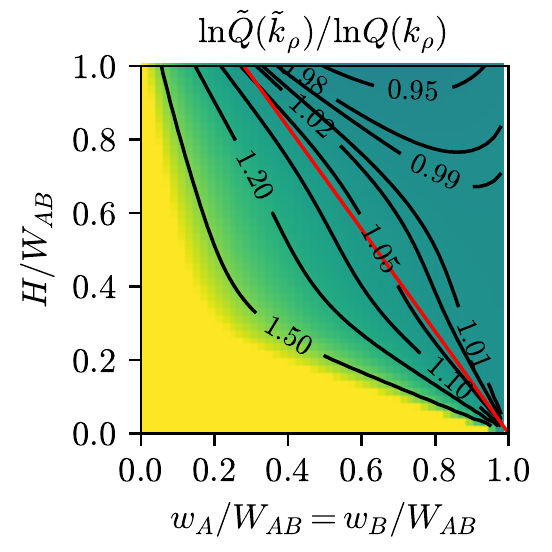}
	\caption{
		Propapation of the ratio of approximation $\ln\tilde{Q}(\tilde{k}_{\rho})/\ln Q(k_{\rho})$ for the case of shallow domains and wide bands.
		Contour lines of selected values are shown in black.
		Line joining the points $(1,0) \to (\num{0,28}, 1)$ is shown in red.
	}
	\label{supp:fig:ratios:shallow:wide}
\end{figure}

This is shown in \fig\ \ref{supp:fig:ratios:tall:narrow} for the case of tall domains and narrow bands.
Unlike the previous case, the first ratio $\ln Q(k_{\rho}')/\ln\tilde{Q}(k_{\rho}') > \num{0,95}$ (relative error of \SI{-5}{\percent}) for widths $w_{E}/W_{AB} < \num{0,82}$ (${k_{\rho}'}^{2} \lesssim \num{0,315}$), which corresponds to a large portion of the parameter domain.
This agrees with the statement in \cite[before \eqn\ (5)]{Morf2006may}, which says that the approximation converges rapidly for gap widths $g > \num{0,1}w_{E} \Leftrightarrow w_{E}/W_{AB} < \num{0,909}$ (no quantitative error is specified).
Unfortunately, the contribution due to the approximation of the complementary modulus $\tilde{k}_{\rho}'$ makes the second ratio $\ln\tilde{Q}(k_{\rho}')/\ln\tilde{Q}(\tilde{k}_{\rho}') < 1$ in most of the domain $(w_{E}/W_{AB}, H/W_{AB})$.
This approximation helps to simplify the expression for the complementary modulus in \eqn\ \eqref{supp:eqn:krho':Hinf} at the expense of shrinking the domain where the total ratio $\ln Q(k_{\rho}')/\ln\tilde{Q}(\tilde{k}_{\rho}') > \num{0,95}$ (relative error of approximation less than \SI{-5}{\percent}), which corresponds to electrode widths $w_{E}/W < \num{0,56}$.

The propagation of the relative error for cases of shallow cells
have not been reported previously in the literature and
they are given below as a reference.

\begin{figure}[t]
	\centering
	\includegraphics{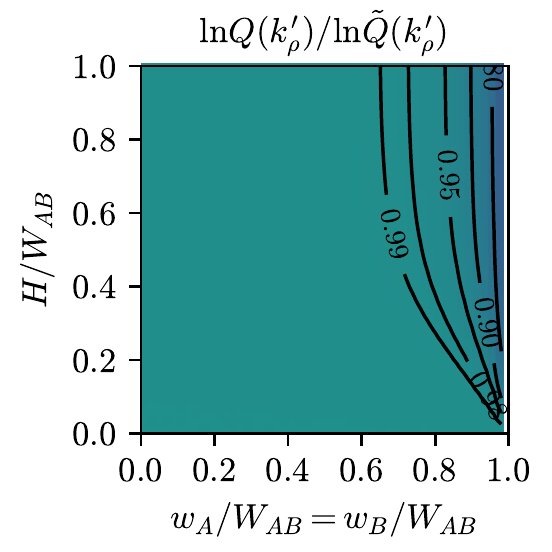} 
	\includegraphics{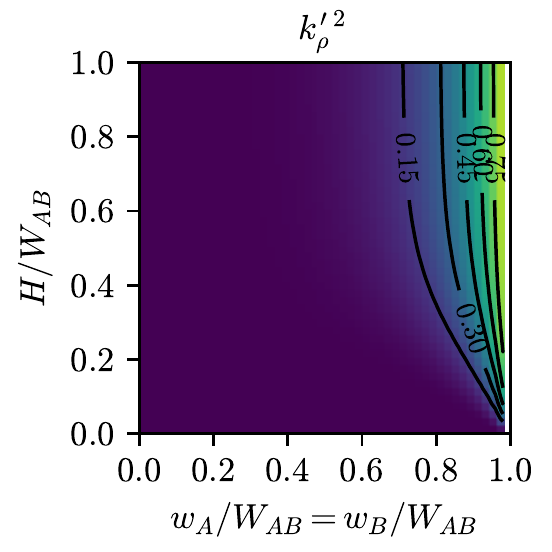}
	\\
	\includegraphics{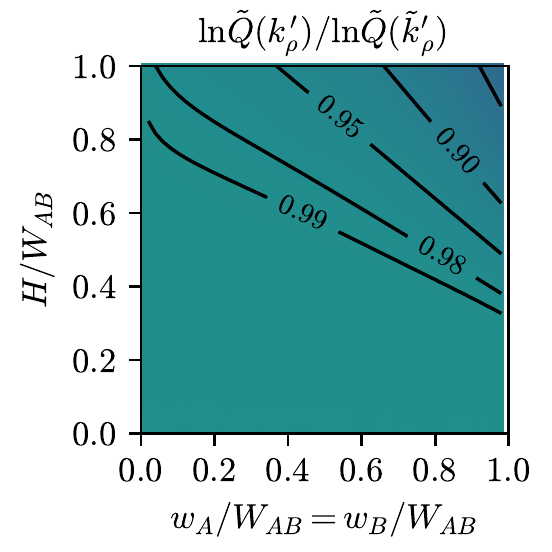} 
	\includegraphics{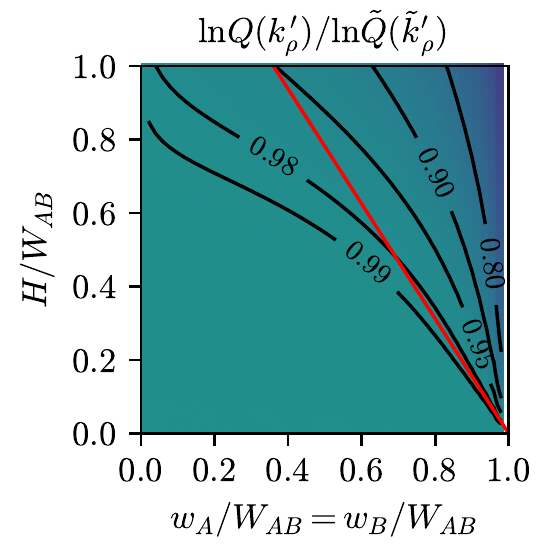}
	\caption{
		Propapation of the ratio of approximation $\ln\tilde{Q}(\tilde{k}_{\rho})/\ln Q(k_{\rho})$ for the case of shallow domains and narrow bands.
		Contour lines of selected values are shown in black.
		Line joining the points $(1,0) \to (\num{0,36}, 1)$ is shown in red.
	}
	\label{supp:fig:ratios:shallow:narrow}
\end{figure}

\fig\ \ref{supp:fig:ratios:shallow:wide} shows the propagation of the total ratio, in \eqn\ \eqref{supp:eqn:ratio:k_rho}, for the case of shallow domains and wide bands.
The first ratio $\ln\tilde{Q}(k_{\rho})/\ln Q(k_{\rho}) < \num{1,05}$ (relative error of \SI{5}{\percent}) for combinations of $(w_{E}/W_{AB}, H/W_{AB})$ roughly to the right of the line joining the points $(1,0) \to (\num{0,94}, 1)$ ($k_{\rho}^{2} \lesssim \num{0,324}$), which again correspond to very wide electrodes.
The contribution due to the approximation of the modulus $\tilde{k}_{\rho}$ in \eqn\ \eqref{supp:eqn:krho:H0} makes the second ratio $\ln\tilde{Q}(\tilde{k}_{\rho})/\ln\tilde{Q}(k_{\rho}) < 1$ in most of the domain $(w_{E}/W_{AB}, H/W_{AB})$.
This also extends the region where the total ratio $\num{0,95} < \ln\tilde{Q}(\tilde{k}_{\rho})/\ln Q(k_{\rho}) < \num{1,05}$ (relative error less than \SI{+-5}{\percent}), which corresponds to combinations of $(w_{E}/W_{AB}, H/W_{AB})$ approximately at the right of the line joining the points $(1,0) \to (\num{0,28}, 1)$.

\fig\ \ref{supp:fig:ratios:shallow:narrow} shows the propagation of the total ratio, in \eqn\ \eqref{supp:eqn:ratio:k'_rho}, for the case of shallow domains and narrow bands.
The first ratio $\ln Q(k_{\rho}')/\ln\tilde{Q}(k_{\rho}') > \num{0,95}$ (relative error of \SI{-5}{\percent}) for combinations of $(w_{E}/W_{AB}, H/W_{AB})$ roughly to the left of the line joining the points $(1,0) \to (\num{0,82}, 1)$ (${k_{\rho}'}^{2} \lesssim \num{0,350}$), which corresponds to a large portion of the parameter domain.
The approximation of the complementary modulus $\tilde{k}_{\rho}'$ makes the second ratio $\ln\tilde{Q}(k_{\rho}')/\ln\tilde{Q}(\tilde{k}_{\rho}') < 1$ in most of the domain $(w_{E}/W_{AB}, H/W_{AB})$.
This approximation helps to simplify the expression for the complementary modulus in \eqn\ \eqref{supp:eqn:krho':H0} at the expense of shrinking the domain where the total ratio $\ln Q(k_{\rho}')/\ln\tilde{Q}(\tilde{k}_{\rho}') > \num{0,95}$ (relative error of approximation less than \SI{-5}{\percent}), which corresponds to combinations of $(w_{E}/W_{AB}, H/W_{AB})$ approximately at the left of the line joining the points $(1,0) \to (\num{0,36}, 1)$.

In summary, the regions of validity for the approximated flux (within \SI{+- 5}{\percent} relative error) correspond to $w_{E}/W_{AB} > \num{0,46}$ (wide bands) and $w_{E}/W_{AB} < \num{0,56}$ (narrow bands) for the case of tall domains $H/W_{AB} > 1$.
In the case of shallow domains $H/W_{AB} \leq 1$, these regions of validity approximately correspond to the right of the line connecting the points $(1,0) \to (\num{0,28},1)$ (wide bands) and to the left of the line connecting connecting the points $(1,0) \to (\num{0,36},1)$ (narrow bands), where the line that passes through the points $(1, 0)$ and $(p, 1)$ is given by
\begin{equation}
	\frac{w_{E}}{W_{AB}} + (1 - p) \frac{H}{W_{AB}} = 1
\end{equation}
that is, $w_{E} + \num{0,72}H \gtrsim W_{AB}$ for wide bands and
$w_{E} + \num{0,64}H \lesssim W_{AB}$ for narrow bands.

\renewcommand{\refname}{\vspace*{-1.5em}}
\section{References}
\bibliographystyle{utphys-cfgy}
\bibliography{refs-suppinfo-urls}

\end{document}